\documentclass[journal]{vgtc}                     

\onlineid{1215}

\vgtccategory{Research}

\vgtcpapertype{algorithm/technique}

\title{{\sysname}: Probabilistic Super Resolution \\ with Flow-Based Models for Scientific Data}

\author{
  \authororcid{Jingyi~Shen}{0000-0001-5478-3993}
  and~\authororcid{Han-Wei~Shen}{0000-0002-1211-2320} 
}

\authorfooter{
  \item Jingyi Shen and Han-Wei Shen are with the Department of Computer Science and Engineering, The Ohio State University. E-mail: \{shen.1250 and shen.94\}@osu.edu.
}

\abstract{ 
Although many deep-learning-based super-resolution approaches have been proposed in recent years, because no ground truth is available in the inference stage, few can quantify the errors and uncertainties of the super-resolved results. 
For scientific visualization applications, however, conveying uncertainties of the results to scientists is crucial to avoid generating misleading or incorrect information. 
In this paper, we propose {\sysname}, a novel normalizing flow-based generative model for scientific data super-resolution that incorporates uncertainty quantification into the super-resolution process. 
{\sysname} learns the conditional distribution of the high-resolution data based on the low-resolution counterpart. By sampling from a Gaussian latent space that captures the missing information in the high-resolution data, one can generate different plausible super-resolution outputs. 
The efficient sampling in the Gaussian latent space allows our model to perform uncertainty quantification for the super-resolved results. 
During model training, we augment the training data with samples across various scales to make the model adaptable to data of different scales, achieving flexible super-resolution for a given input. 
Our results demonstrate superior performance and robust uncertainty quantification compared with existing methods such as interpolation and GAN-based super-resolution networks.
} 


\newcommand{\clr}{\textcolor{black}} 
 
\newcommand{\sysname}{PSRFlow} 

\graphicspath{{figs/}{figures/}{pictures/}{images/}{./}} 

\usepackage{tabu}                      
\usepackage{booktabs}                  
\usepackage{lipsum}                    
\usepackage{mwe}                       

\usepackage{mathptmx}                  
\usepackage[ruled,vlined]{algorithm2e}
\usepackage{fontawesome}
\usepackage{amsfonts}
\usepackage{multirow}
\usepackage{makecell}
\usepackage{amsmath}

\begin{document}


\maketitle


\section{Introduction}
The ability to conduct high-resolution simulations is crucial for obtaining a deep understanding of scientific phenomena. When data are stored at a lower resolution due to I/O and storage constraints, the resulting low-quality visualization suffers from information loss, which makes scientific discovery more difficult. Recently, many deep learning-based methods have been proposed for scientific data super-resolution. These methods learn the complex correspondence between low and high-resolution data, and have shown remarkable performance. \cite{An2021STSRNet, Weiss2021isoSR, Han2022SSR-TVD, Guo2020SSR-VFD, Han2020TSR-TVD, Wurster2022HierSR, han2021stnet, xie2018tempogan}

Although widely used, several limitations still remain for deep learning-based super-resolution methods.
First, current methods learn a deterministic one-to-one mapping between high and low-resolution data pairs. During inference, given a low-resolution input, the model predicts one single high-resolution output. However, due to the loss of information in the low-resolution data, it is not possible to precisely determine the exact high-resolution output given the input, because one low-resolution data may correspond to multiple high-resolution outputs. Existing methods do not directly tackle the challenge of modeling such variations in the super-resolution process. 
Another challenge of the existing learning-based super-resolution methods is that in the inference stage, since no high-resolution ground truth is available, quantifying the errors and uncertainties of the super-resolved result is difficult. 
However, to avoid making wrong decisions based on the super-resolved data, it is crucial to convey uncertainties and provide indications about the quality of the super-resolution outputs to scientists. 
Currently, there is relatively little research on modeling the uncertainties associated with neural network-based super-resolution outputs.



In this paper, we propose {\sysname}, a novel probabilistic super-resolution approach based on the invertible normalizing flow~\cite{kingma2018glow, Dinh2017RealNVP, rezende2015variational}.
Normalizing flows are constructed based on a sequence of invertible transformations that convert a simple distribution into a more complex one. By using a conditional normalizing flow to model the distribution of high-resolution data conditioned on low-resolution inputs, we can overcome the limitations of existing techniques. 
First, rather than modeling a deterministic one-to-one mapping, we treat super-resolution as a conditional generation process. We model the prior knowledge of missing high-frequency information as a Gaussian distribution in the latent space conditioned on the low-resolution data. We choose Gaussian distributions since they are easy to work with and efficient to sample. Once the conditional distribution is well-learned, we can sample a latent vector from the distribution and reconstruct plausible high-resolution data. 
Second, we utilize normalizing flows to capture the complex relationship between low and high-resolution data. By modeling the missing information of high-resolution data as simpler Gaussian distributions in the latent space, {\sysname} can explicitly capture the variations among the high-resolution data for a given low-resolution input. Samples drawn from the Gaussian latent space reflect the uncertainties in the high-resolution data generation process. Regions with higher variations indicate that the model is less confident in those areas and thus errors are more likely to occur. As a result, the reliability of {\sysname} is boosted, since even without the ground truth it is still possible to convey potential errors to scientists during inference.

Our {\sysname} is an invertible framework that effectively captures the relationships between low and high-resolution data. 
The forward process of {\sysname} gradually transforms the high-resolution data in the original complex data space to a latent space with two components. One latent subspace encodes low-resolution information, while the other latent subspace follows a Gaussian distribution and captures the missing high-frequency information conditioned on the low-resolution information. 
In the reverse process of {\sysname}, \clr{missing high-frequency information can be estimated by sampling from the Gaussian latent space.} By combining this information with the low-resolution latent vector extracted by an encoder, and leveraging the invertible transformations of normalizing flow, {\sysname} can produce a high-quality super-resolution output. When sampling the Gaussian latent space multiple times, scientists can estimate the uncertainties of the super-resolution results by computing the variations of the super-resolved outputs at each voxel. 
In addition, during training, we augment the training data with samples from various scales so that once {\sysname} is trained, it can be adaptively applicable for super-resolution across different scales, allowing us to achieve flexible super-resolution for a given low-resolution input with one trained model. 

Based on the learned distribution mappings, given low-resolution data, scientists can obtain different plausible super-resolved results by sampling the Gaussian latent space. In contrast to other super-resolution methods in the visualization literature, our method allows exploring the super-resolved data space and estimating the uncertainties of the super-resolution results. 
The results of our qualitative and quantitative evaluations demonstrate that our super-resolution method has superior performance with robust uncertainty quantification, compared to existing interpolation and generative adversarial network (GAN) based super-resolution methods. 
The contributions of our work are threefold:  
\vspace{-10pt}
\begin{itemize}
\item First, we propose a novel probabilistic super-resolution approach based on normalizing flow to model the missing information of high-resolution data as distributions instead of a deterministic one-to-one mapping as in the existing learning-based methods.
\vspace{-4pt}
\item Second, our method provides uncertainty quantification to increase the reliability of the model for scientists to make informed decisions.
\vspace{-4pt}
\item Third, with data augmentation and cross-scale training, we allow flexible super-resolution for data from different resolution levels with one trained model.
\end{itemize}

\vspace{-3pt}

\section{Related Works}
We adopt invertible normalizing flow for probabilistic super-resolution with uncertainty estimation. In this section, we summarize the related works of scientific data super-resolution, uncertainty analysis of scientific data, and the applications of normalizing flow. 

\textit{Scientific Super-Resolution for Visualization.}
Many deep-learning-based super-resolution approaches have been proposed for scientific data. SSR-TVD\cite{Han2022SSR-TVD} utilizes a generative adversarial network (GAN) for spatial super-resolution of time-varying data with temporal coherence. SSR-VFD\cite{Guo2020SSR-VFD} upscales the low-resolution vector field spatially. 
Weiss et al. \cite{Weiss2021isoSR} upscale low-resolution isosurfaces to high-resolution with ambient occlusion to reduce full-resolution rendering time. Instead of uniform grid super-resolution, Wurster et al. \cite{Guo2020SSR-VFD} utilize a hierarchy of super-resolution models for upscaling data with varying levels of detail. 
For temporal super-resolution, TSR-TVD~\cite{Han2020TSR-TVD} generates missing time steps in time-varying data from a low temporal resolution volume sequence. 
STSRNet~\cite{An2021STSRNet} introduces a two-stage framework (upscale temporal dimension first then spatial) for spatial-temporal super-resolution of time-varying vector fields given low-resolution key time steps as input. 
STNet~\cite{han2021stnet} utilizes feature-space temporal interpolation and feature upscaling to synthesize spatiotemporal super-resolution volumes for time-varying data. TempoGAN~\cite{xie2018tempogan} utilizes a novel temporal discriminator to produce highly detailed and temporally coherent four-dimensional fields for fluid flows. 
However, no existing work has investigated the uncertainties introduced in the super-resolution process. For scientific visualization and analysis, conveying uncertainties is crucial for scientists to avoid making misleading or incorrect decisions. 
To fill this gap, in this paper, we propose probabilistic super-resolution with uncertainty quantification for scientific visualization.

\textit{Normalizing Flow Applications.}
Our work is based on one of the powerful generative models named normalizing flow~\cite{kingma2018glow, Dinh2017RealNVP, rezende2015variational}. Due to its ease of training and fast inference, normalizing flow has been used for various tasks such as image super-resolution~\cite{lugmayr2020srflow, Jo2021TacklingIllPosed, jo2021srflow-da, liang2021hierarchical}, point cloud generation~\cite{Yang2019PointFlow, Postels2021MixNFPoint}, and speech synthesis~\cite{Aggarwal2020NFTextToSpeech}, etc. 
Our work leverages the theory presented in Ardizzone et al.~\cite{ardizzone2018analyzing}, which approximates a conditional distribution through a tractable distribution where an extra latent variable is introduced to mitigate the information loss. In our case, the lost information is the high-frequency details that get smoothed out during downsampling which can be captured by the Gaussian latent variable. 

\textit{Uncertainty Visualization.}
Uncertainty visualization has become a top visualization research direction to avoid generating misleading interpretations of data. Uncertainties can be introduced in different stages of visualization pipeline, including data acquisition~\cite{Vietinghoff2022CPProb, pothkow2011probabilisticMC, pothkow2010positional, athawale2022fiber, athawale2020uncertainty}, data transformation~\cite{Correa2009UQAVA, schlegel2012interpolationUQ, Hägele2023UAMS}, visual mapping and rendering~\cite{liu2016uncertainty, pang1997approaches2UV, Grigoryan2004PointSurfaceUQ, fout2012fuzzyVR, Sakhaee2017StatisticalDVR, athawale2020direct} stages. In deep learning for scientific visualization, Han et al.~\cite{han2022exploratory} found that the model's performance is affected by flow behavior. Regions with greater separation in the flow field have higher errors. 
Our work focuses on the uncertainty in the data transformation stage, where low-resolution data are super-resolved into high-resolution counterparts. 

 \section{Background: Normalizing Flow Model}\label{sect:BG_NF}
Our work is based on a conditional normalizing flow to model the complex relationship between low and high-resolution data for scientific super-resolution with uncertainty estimation. In this section, we briefly introduce the normalizing flow model. 
 
Normalizing flows is a type of generative neural network used to model complex probability distributions. It learns a bijective mapping between the data space with a complex distribution and a latent space with a simple base distribution (typically a Gaussian distribution) through a sequence of simple and invertible mappings (i.e., ``flows''). By learning such bijective mappings, one can model the complex probability density of data and sample from the simple latent distribution to produce in-distribution complex data samples.

\begin{figure}[htp]
    \centering
    \includegraphics[width=\columnwidth]{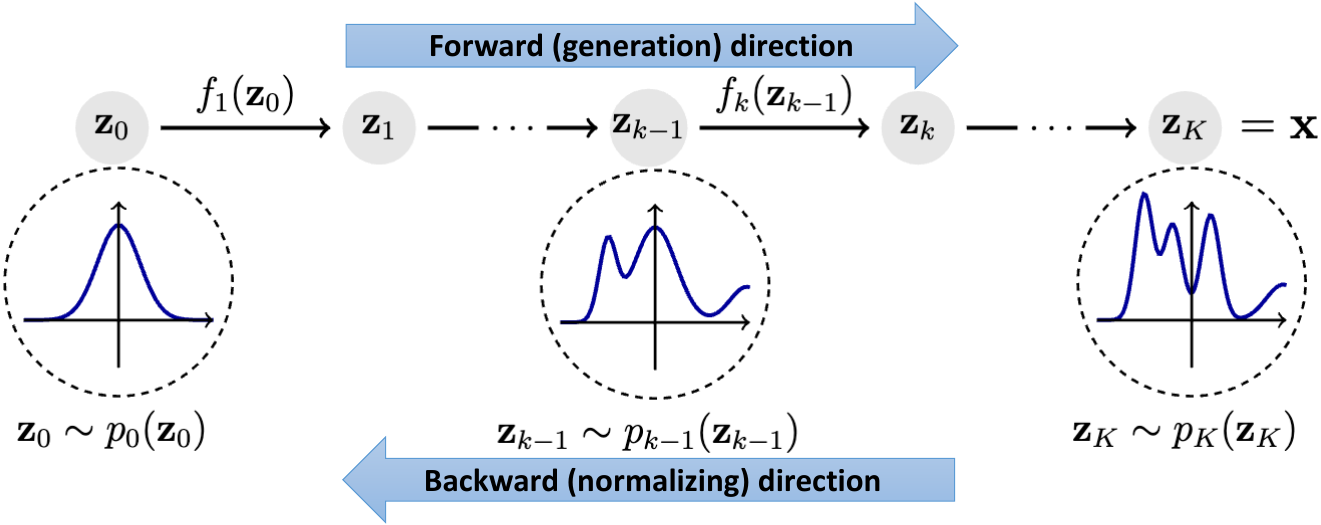}
    \vspace{-18pt}
    \caption{\clr{Normalizing flow distribution transformation}~\cite{Janosh2022TikZNF}.
    The forward direction transforms a simple distribution into a complex one step by step. This process is invertible in the backward direction.}
    \label{fig:bg_nf}
    \vspace{-2pt}
\end{figure}

As shown in~\cref{fig:bg_nf}, if we denote the invertible mapping as $f$, the \textit{\textbf{forward direction}} (or \textit{\textbf{generation direction}}) of $f$ is to gradually transform a variable $Z$ with a known and tractable probability density \clr{$p(Z)$} into a random variable $X$ with complex probability density \clr{$p(X)$}, denoted as $X=f(Z)$. 
Note that one can generate a sample $\mathbf{x}$ by sampling $\mathbf{z}$ from the base distribution $p(Z)$ and applying the forward transformation $\mathbf{x}=f(\mathbf{z})$.
Due to the invertibility of $f$, the \textit{\textbf{backward direction}} (or \textit{\textbf{normalizing direction}}) of $f$ is to transform from complex data distribution into simple latent distribution: $Z=f^{-1}(X)$.

\vspace{-12pt}

\noindent Generally, the invertible function $f$ is composed of $K$ invertible and learnable transformations: $f=f_1 \circ f_2 \circ \cdot \cdot \cdot \circ f_K$. So we have: 
\vspace{-2pt}
\begin{equation} \label{eq:nf}
\mathbf{x} = \mathbf{z}_K = f_{K} \circ f_{K-1} \circ \cdot \cdot \cdot \circ f_{1}(\mathbf{z}_0)
\end{equation}
\noindent where $\mathbf{z}_0$ has a probability density $p_0(\mathbf{z}_0)$, which is the base distribution (e.g. a standard normal distribution). Each $f_k$ is an invertible and differentiable transformation. 
The probability density of $\mathbf{x}$ is computed by repeatedly applying the change-of-variable formula, and since the determinant of products equals the product of determinants, we have:
\vspace{-4pt}
\begin{equation} \label{eq:nf_px}
p(\mathbf{x}) = p_K(\mathbf{z}_K) = p_0(\mathbf{z}_0)\prod_{k=1}^{K} \left| \det{\frac{\partial f_k(\mathbf{z}_{k-1})} {\partial \mathbf{z}_{k-1}}} \right|^{-1} 
\vspace{-2pt}
\end{equation}
Taking logarithm of \cref{eq:nf_px}, we have:
\vspace{-4pt}
\begin{equation} \label{eq:nf_logpx}
\log p(\mathbf{x}) = \log p_K(\mathbf{z}_K) = \log p_0(\mathbf{z}_0) - \sum_{k=1}^{K} \log \left| \det{\frac{\partial f_k(\mathbf{z}_{k-1})} {\partial \mathbf{z}_{k-1}}} \right| 
\vspace{-3pt}
\end{equation}
\noindent In \cref{eq:nf_logpx}, the first term on the right is the log probability density of the latent distribution, and the second term is the volume change caused by the transformation. It has been formally proven that if the transformation function $f$ can be arbitrarily complex, one can generate any target distribution with any latent distribution~\cite{Bogachev2005TriangularTO}. 
Function $f$ is parameterized by $\theta$ and is learnable. The optimization goal for normalizing flows is to find $\theta$ that maximizes the log-likelihood, which is equal to $\log p(\mathbf{x})$ and can be directly computed using \cref{eq:nf_logpx}. 
One way to interpret this is from the statistical distance perspective, where maximizing log-likelihood is equivalent to minimizing the KL divergence between true data distribution and the flow-parameterized data distribution. 

\begin{figure}[htp]
    \centering
    \includegraphics[width=6.5cm]{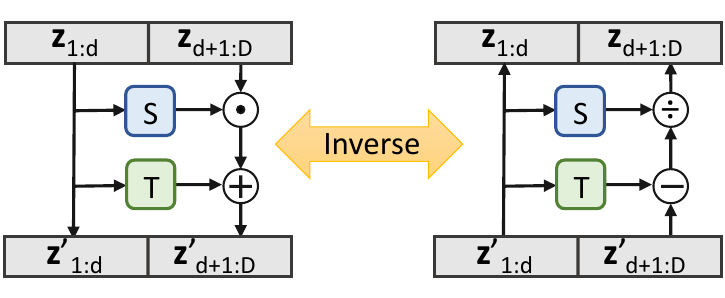}
    \vspace{-10pt}
    \caption{\clr{The computation graph of an affine coupling layer in the forward (left) and its inverse (right) direction.}}
    \label{fig:bg_coupling}
\vspace{-10pt}
\end{figure}

Constructing invertible transformations can be difficult since it requires invertibility of each $f_k$ and easy-to-compute Jacobian determinant $det(J_{f_k})$ in~\cref{eq:nf_logpx}. 
One common approach is utilizing the invertible ``affine coupling layer''~\cite{Dinh2017RealNVP} whose determinant can be efficiently computed as the product of diagonal elements of the Jacobian matrix. 
As shown in \cref{fig:bg_coupling} (left), the affine coupling layer partitions the input $\mathbf{z}$ of this layer into two subspaces ($\mathbf{z}_{1:d}$ and $\mathbf{z}_{d+1:D}$) and applies a differentiable affine transformation to $\mathbf{z}_{d+1:D}$ while keeping $\mathbf{z}_{1:d}$ untouched. $d$ is the splitting dimension. The scale and shift parameters for affine transformation are computed based on the untouched $\mathbf{z}_{1:d}$. Formally, the affine coupling layer's output $\mathbf{z}^\prime$ can be computed by:
\vspace{-5pt}
\begin{equation}\label{eq:affine_coupling}
\begin{gathered} 
\mathbf{z}^\prime_{1:d} = \mathbf{z}_{1:d} \\
\mathbf{z}^\prime_{d+1:D} = \mathbf{z}_{d+1:D} \odot \exp(S(\mathbf{z}_{1:d})) + T(\mathbf{z}_{1:d})
\end{gathered}
\vspace{-4pt}
\end{equation}
\noindent $\odot$ denotes the element-wise Hadamard product. Functions $S(\cdot)$ and $T(\cdot)$ can be highly complex (e.g., neural networks) and non-invertible. They take $\mathbf{z}_{1:d}$ as input and predict scale and translation parameters to apply on $\mathbf{z}_{d+1:D}$. The high modeling capability of $S(\cdot)$ and $T(\cdot)$ is the key to the effectiveness of coupling layers.  
The inverse of \cref{eq:affine_coupling} is:
\vspace{-5pt}
\begin{equation}\label{eq:affine_coupling_inverse}
\begin{gathered} 
\mathbf{z}_{1:d} = \mathbf{z}^\prime_{1:d} \\
\mathbf{z}_{d+1:D} = (\mathbf{z}^\prime_{d+1:D} - T(\mathbf{z}^\prime_{1:d})) \odot \exp(-S(\mathbf{z}^\prime_{1:d}))
\end{gathered}
\end{equation}
\vspace{-19pt}

Note that one subspace of the variable (i.e., $\mathbf{z}_{1:d}$) remains identical after transformation, which reduces the modeling ability of the normalizing flow. To solve this, channel order permutation \cite{Dinh2017RealNVP, kingma2018glow} is usually adopted to make sure all variable dimensions got transformed.
Besides this, normalizing flows often contain invertible activation normalization layers (i.e., ``actnorm'') \cite{kingma2018glow} to normalize feature maps for stable gradients and faster convergence. 
By stacking multiple invertible coupling layers along with normalization and permutation (known as flow steps), normalizing flows can scale and shift input variables multiple times and learn complex dependencies between variables, making it possible to transform between complex and simple distributions.

As a generative model, normalizing flows do not have mode collapse and training instability issues like generative adversarial networks (GANs) or Variational Autoencoders (VAEs)~\cite{arjovsky2017GAN}. Meanwhile, normalizing flows allow exact log-likelihood evaluation and efficient latent space sampling. With this, we can use normalizing flows to model complex distribution mappings between low and high-resolution data. 
In the next section, we will present our uncertainty-aware super-resolution method for scientific data based on a conditional normalizing flow.  




\section{Method}
Super-resolution is a challenging problem due to the inherent difficulties in recovering missing information. In our paper, instead of learning a deterministic one-to-one mapping between low and high-resolution data, we propose {\sysname}, which takes a probabilistic approach to model the conditional distribution of missing information in high-resolution data given the low-resolution input. 
Our approach not only allows high-quality super-resolution but also provides uncertainty estimation of the results, which is particularly important for scientists to make reliable decisions about the underlying scientific phenomenon. 
\subsection{Overview}
\begin{figure}[htp]
    \centering
    \includegraphics[width=8.5cm]{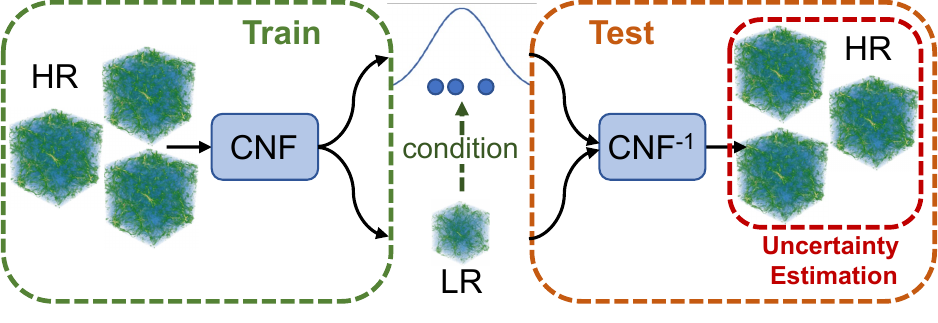}
    \vspace{-11pt}
    \caption{Overview of {\sysname}. 
    High-frequency details and low-resolution (LR) information are modeled separately in the latent space through a conditional normalizing flow (CNF). The high-frequency latent follows a distribution conditioned on LR. \clr{During testing, given LR input, one can sample from the conditional distribution and utilize the inverse of CNF to obtain HR outputs. HR outputs are then used for uncertainty estimation.}}
    \label{fig:overview}
    \vspace{-14pt}
\end{figure}

Given the low-resolution data $Y \in\mathbb{R}^{D \times H\times W}$, we aim to generate its super-resolved counterpart $X \in\mathbb{R}^{rD\times rH\times rW}$, where $r$ is the upscaling factor. \Cref{fig:overview} shows the overview of {\sysname}. 
The motivation for using the normalizing flow model for our task is threefold. 
First, instead of a deterministic one-to-one mapping, normalizing flow allows explicit and efficient probabilistic modeling of the relationship between high and low-resolution data. Second, compared to other generative models like GANs and VAEs, normalizing flows are more stable. Third, straightforward latent space sampling of normalizing flow allows uncertainty quantification of the super-revolved outputs.
{\sysname} utilizes a conditional normalizing flow (CNF) to convert high-resolution data into a latent space consisting of two parts. One part contains low-resolution information, while the other is a Gaussian latent space that encodes the missing high-frequency information conditioned on the low-resolution data. 
The reason for transforming data into a latent space with Gaussian distribution is that Gaussian space is easier to control and sample. During inference, one can sample from the Gaussian latent space for high-frequency information reconstruction. \clr{By combining the sampled high-frequency details with the low-resolution information} and utilizing the inverse transformation of the conditional normalizing flow ($\mathrm{CNF}^{-1}$), we can obtain a possible high-resolution output.

\subsection{Conditional Normalizing Flow for Super Resolution}
The general procedure of super-resolution is to recover high-resolution data $X$ by predicting the missing high-frequency details of a given low-resolution data $Y$. 
While existing super-resolution approaches learn a deterministic mapping $Y \rightarrow X$, we focus on modeling the conditional distribution $P(X \mid Y)$. In this section, we discuss: (1) how to capture the complex relationship between low and high-resolution data, and (2) how to model the missing high-frequency information. 

\subsubsection{High-Resolution Conditional Modeling}\label{sect:high-res-cond}
\clr{Modeling distribution $P(X \mid Y)$ is more challenging than predicting one high-resolution output, as additional effort is required to capture the complex information such as how the data is placed in high-dimensional space.}
To solve this, we propose a data-driven approach by utilizing an invertible conditional normalizing flow to parameterize and learn such conditional distribution. 

Inspired by Fourier domain signal processing, one can decompose data into high and low-frequency components. Given a smooth low-resolution input, it is possible to get a high-resolution output by adding high-frequency details into the low-resolution data.
With this, based on the probability chain rule, the conditional distribution $P(X \mid Y)$ can be further explained into:
\vspace{-3pt}
\begin{equation} \label{eq:P(X|Y)}
P(X \mid Y) = P(X_h, X_l \mid Y) = P(X_h \mid X_l,Y)P(X_l \mid Y) 
\end{equation} 

\vspace{-3pt}
\noindent where $X_h$ and $X_l$ are the high and low-frequency components, respectively. $P(X_h \mid X_l,Y)$ means instead of random noise, we consider the missing high-frequency details to be conditioned on the low-resolution information.
However, modeling distribution in the original data space is non-trivial due to the intractable probability density. This motivates us to utilize conditional normalizing flow since it learns to transform data with complicated and intractable probability density $P(X|Y)$ into a simpler latent space with tractable probability density $P(Z|Y)$. If the model is well-trained, we can derive $P(X|Y)$ explicitly from $P(Z|Y)$ via~\cref{eq:nf_px}. Transforming data into the latent space can also facilitate information decomposition of the high-resolution data. 

\begin{figure}[htp]
    \centering
    \includegraphics[width=\columnwidth]{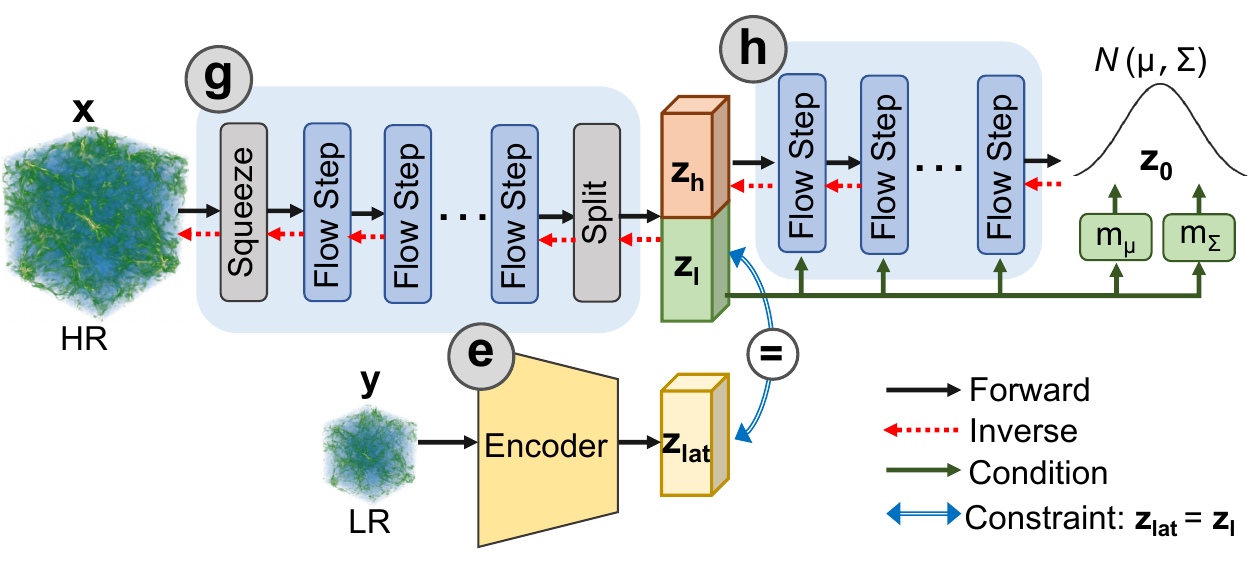}
    \vspace{-20pt}
    \caption{\clr{Architecture of {\sysname}}. During training, input high-resolution $\mathbf{x}$ is transformed into a latent space with two components: $\mathbf{z}_h$ (high-frequency information) and $\mathbf{z}_l$ (low-resolution information) by flow $g$. Encoder $e$ is used to facilitate information decomposition \clr{so that $\mathbf{z}_l$ matches the target low-resolution latent representation $\mathbf{z}_{lat}$ extracted by $e$.} 
    $\mathbf{z}_h$ is further transformed by conditional flow $h$ into $\mathbf{z}_0$ with a Gaussian distribution. During inference, encoder $e$ and the inverse of $g$ and $h$ (red dashed arrow) are used to recover a high-resolution output.} 
    \label{fig:architecture}
\vspace{-4pt}
\end{figure}

As shown in \cref{fig:architecture}, given a pair of high-resolution $\mathbf{x} \in X$ and low-resolution data $\mathbf{y} \in Y$, we first transform $\mathbf{x}$ through a sequence of normalizing flow steps into a latent representation $\mathbf{z}$. Then, we decompose $\mathbf{z}$ into two components, namely $\mathbf{z}_{h}$ and $\mathbf{z}_{l}$, where we force $\mathbf{z}_{h}$ to capture high-frequency details and $\mathbf{z}_{l}$ to capture the low-frequency information based on the low-resolution data $\mathbf{y}$. We formulate this process as:
\vspace{-3pt}
\begin{equation} \label{eq:g(x)}
g(\mathbf{x}) = \mathbf{z} \textrm{,} \quad \textrm{where} \ \mathbf{z}=(\mathbf{z}_{h}, \mathbf{z}_{l})
\vspace{-3pt}
\end{equation}
\begin{equation} \label{eq:p(zhzl|y)}
p(\mathbf{z}_{h}, \mathbf{z}_{l}\mid\mathbf{y}) = p(\mathbf{z}_{h}\mid\mathbf{z}_{l},\mathbf{y})p(\mathbf{z}_{l}\mid\mathbf{y})
\end{equation}

\noindent where $g$ is the invertible normalizing flow and $(\textrm{,})$ means $\mathbf{z}$ is decomposed into $\mathbf{z}_{h}$ and $\mathbf{z}_{l}$. The decomposition in {\sysname} is implemented by latent space channel splitting. This splitting can be interpreted as projecting information onto two subspaces of the latent space.
The inverse of $g$ is the generative direction: $g^{-1}((\mathbf{z}_{h}, \mathbf{z}_{l})) = \mathbf{x}$. 
\Cref{eq:p(zhzl|y)} is the key and goal for our probabilistic super-resolution. The only difference to~\cref{eq:P(X|Y)} is that this is modeled in the latent space but will be transformed back to the data space with the help of flow $g$.

\vspace{-12pt}

The challenge now is how to enforce $\mathbf{z}_{h}$ and $\mathbf{z}_{l}$ to capture the desired information.
In many normalizing flow applications~\cite{ardizzone2018analyzing,Yang2019PointFlow}, input data is transformed into a standard Gaussian distribution in the latent space, i.e., $\mathbf{z} \sim \mathcal{N}(0, I)$. However, scientific data often have complex features across different regions and time steps, and representing such complicated information using a standard Gaussian distribution may not be feasible. 
Thus, we impose a constraint on $\mathbf{z}_{l}$ to enforce it to encode low-resolution information, whereas we further transform $\mathbf{z}_{h}$ into a Gaussian latent space that captures the missing information between low and high-resolution data. Instead of standard Gaussian distributions, the parameters of this Gaussian latent space are predicted based on the low-resolution data. 

\subsubsection{Low-Resolution Encoding}
To ensure that $\mathbf{z}_{l}$ encodes the low-resolution information, in \cref{fig:architecture}, {\sysname} has another low-resolution encoding network $e$ to add explicit supervision to $\mathbf{z}_{l}$.
Encoder $e$ does not need to be invertible, and is built based on the popular ResNet (Residual Network)~\cite{He2016ResNet} to progressively extract a low-resolution representation $\mathbf{z}_{lat}$ from low-resolution data $\mathbf{y}$: 
\vspace{-4pt} 
\begin{equation} \label{eq:z_lat}
\mathbf{z}_{lat}=e(\mathbf{y})
\vspace{-4pt}
\end{equation}
During training, $\mathbf{z}_{l}$ is constrained to match $\mathbf{z}_{lat}$. During inference, with the well-trained encoder $e$, given a low-resolution data $\mathbf{y}$, we can replace $\mathbf{z}_{l}$ with encoder generated $\mathbf{z}_{lat}$ without hurting the reconstruction quality of normalizing flow $g$. This is ensured by the invertibility of flow $g$. So we have $g^{-1}((\mathbf{z}_{h}, \mathbf{z}_{l})) = g^{-1}((\mathbf{z}_{h}, \mathbf{z}_{lat})) = \mathbf{x}$. 
Due to the deterministic mapping in~\cref{eq:z_lat} and since $\mathbf{z}_{l}$ is forced to match $\mathbf{z}_{lat}$, we have $p(\mathbf{z}_{lat}\mid\mathbf{y})=p(\mathbf{z}_{l}\mid\mathbf{y})=1$. We reformulate~\cref{eq:p(zhzl|y)} into:
\vspace{-3pt} 
\begin{equation} \label{eq:p(zhzl|y)_zl}
p(\mathbf{z}_{h}, \mathbf{z}_{l}\mid\mathbf{y}) =p(\mathbf{z}_{h}\mid\mathbf{z}_{l},\mathbf{y})= p(\mathbf{z}_{h}\mid\mathbf{y})
\end{equation}
\Cref{eq:p(zhzl|y)_zl} indicates that instead of treating the decomposed high-frequency information as independent noise, we consider it to be dependent on the low-resolution data. 

\subsubsection{High-Frequency Modeling}
To fully capture the complex variations of high-frequency details given the low-resolution data, i.e., $p(\mathbf{z}_{h}\mid\mathbf{y})$, in~\cref{fig:architecture}, we further transform $\mathbf{z}_{h}$ through a conditional normalizing flow $h$ into a simpler Gaussian latent space $\mathbf{z}_{0}$ with predicted mean $\mu$ and variance $\Sigma$. Different from previous layers in unconditional flow $g$, flow steps in $h$ take an additional input $\mathbf{z}_{l}$ as a condition when transforming its input $\mathbf{z}_{h}$ into the innermost latent space $\mathbf{z}_{0}$, as shown in~\cref{fig:architecture} (green arrow). The condition $\mathbf{z}_{l}$ is used as additional information for scaling and transformation parameter prediction of affine coupling layers in flow $h$. More details about affine coupling layers can be found in~\cref{sect:BG_NF}. Thus, we have:
\vspace{-5pt} 
\begin{equation} \label{eq:g_a}
h(\mathbf{z}_{h}, \mathbf{z}_{l})=\mathbf{z}_0
\vspace{-2pt}
\end{equation}
\begin{equation} \label{eq:p(z0|y)}
p(\mathbf{z}_0\mid\mathbf{y}) = \mathcal{N}(\mu,\Sigma)\textrm{,} \quad \textrm{where} \ {\mu}=m_{\mu}(\mathbf{y}) \ \textrm{and} \ {\Sigma}=m_{\Sigma}(\mathbf{y})
\end{equation}
where $m_{\mu}$ and $m_{\Sigma}$ are Gaussian distribution prediction networks for the missing information. Due to the invertibility of flow $h$, we can derive $p(\mathbf{z}_{h}\mid\mathbf{y})$ from Gaussian distribution $p(\mathbf{z}_0\mid\mathbf{y})$. Further, based on the invertibility of flow $g$, we can derive $p(\mathbf{x}\mid\mathbf{y})$ from $p(\mathbf{z}_{h},\mathbf{z}_{l}\mid\mathbf{y})$. 

\subsection{Uncertainty Quantification}\label{sect:UQ}
Uncertainty quantification is crucial for scientific visualization and analysis of super-resolved data since it will provide scientists with indications about the quality of results. In this section, we discuss how to estimate uncertainties of {\sysname}'s high-resolution outputs. 
\vspace{-1pt}
\subsubsection{Conditional Distribution Modeling}
It is intractable to compute the conditional distribution of high-resolution data $\mathbf{x}$ given low-resolution data $\mathbf{y}$ in the original data space. As discussed in the previous section, we parameterize such conditional distribution through the proposed {\sysname}. According to \cref{eq:nf_px}, \cref{eq:p(zhzl|y)_zl} and \cref{eq:p(z0|y)}, exact conditional distribution $p(\mathbf{x}\mid\mathbf{y})$ can be explicitly computed via the generation direction of flow as: 
\vspace{-8pt}
\begin{align} \label{eq:p(x|y)}
\begin{gathered}
p(\mathbf{x}\mid \mathbf{y}) = p(\mathbf{z}_{h}\mid \mathbf{y})\prod_{k=1}^{K_g} \left| \det{\frac{\partial g^{-1}_{k}(\mathbf{z}_{k-1})} {\partial\mathbf{z}_{k-1}}} \right|^{-1} , \\
\textrm{where} \ p(\mathbf{z}_{h}\mid \mathbf{y}) = p(\mathbf{z}_0\mid \mathbf{y})\prod_{k=1}^{K_h} \left| \det{\frac{\partial h^{-1}_{k}(\mathbf{z}_{k-1})} {\partial\mathbf{z}_{k-1}}} \right|^{-1} \\
\textrm{and} \ p(\mathbf{z}_0\mid \mathbf{y}) = \frac{1}{\sqrt{(2\pi)^{D} |\Sigma|}} \exp\left(-\frac{1}{2} (\mathbf{z}_0 - {\mu})^\top \Sigma^{-1} (\mathbf{z}_0 - {\mu})\right)
\vspace{-4pt}
\end{gathered} 
\end{align}
\noindent where $\mathbf{z}_{h}$ is the high-frequency latent component and $D$ is the dimensionality of the Gaussian latent  $\mathbf{z}_0$.
${K_g}$ and ${K_h}$ are the total number of invertible flow steps in $g$ and $h$, respectively. $g^{-1}_{k}$ is inverse of the k-th layer in flow $g$ and $\mathbf{z}_{k-1}$ is the input of this inverse function. 

As discussed in~\cref{sect:BG_NF}, maximizing the conditional log-likelihood $\log p(\mathbf{x}\mid\mathbf{y})$ is equal to minimizing the KL divergence between the true conditional data distribution and the flow parameterized data distribution. Thus, we incorporate $\log p(\mathbf{x}\mid\mathbf{y})$ into our training loss in~\cref{sect:lossFunction}. Once the model is well-trained, high-frequency details will be captured in the Gaussian latent space $\mathbf{z}_0$. One advantage of {\sysname} is that it allows accurate and efficient posterior sampling of high-resolution data by sampling from this Gaussian latent space. 

\subsubsection{Uncertainty Aware Super-Resolution}
With explicit conditional distribution modeling, we can efficiently sample the Gaussian latent space for high-quality high-resolution data reconstruction. Moreover, these samples reflect the uncertainty in the high-resolution data generation process. In this section, we discuss how to quantify uncertainties in the super-resolution process.

During inference, to produce one high-resolution output, as shown in~\cref{alg:uq}, {\sysname} first extracts the low-resolution latent representation $\mathbf{z}_{lat}$ from the low-resolution input $\mathbf{y}$ using encoder $e$. Due to the constraint that we added during training, $\mathbf{z}_{lat}$ can replace $\mathbf{z}_{l}$ without hurting the representation quality. 
Then, {\sysname} utilizes $\mathbf{z}_{l}$ (equal to $\mathbf{z}_{lat}$) to predict mean ($\mu$) and variance($\Sigma$) of the innermost Gaussian latent space $\mathbf{z}_{0}$. This latent space captures the variations of the missing high-frequency details and by sampling from it and utilizing the inverse of flow $h$, {\sysname} can recover a plausible missing high-frequency information $\mathbf{z}_{h}$. {\sysname} then concatenates $\mathbf{z}_{h}$ and $\mathbf{z}_{l}$, and apply the inverse of flow $g$ to obtain one possible high-resolution output. 
By sampling the Gaussian latent space and repeating this process by $N$ times, let $\mathbf{x}^{\prime(i)}$ denote the $i$-th generated high-resolution output, we can compute the mean $\bar{\mathbf{x}}$ and standard deviation $\mathbf{x}_{\sigma}$ of high-resolution outputs using \cref{alg:uq}. The mean $\bar{\mathbf{x}}$ is used as the super-resolution final output and the 
standard deviation $\mathbf{x}_{\sigma}$, which measures the variations in the high-resolution data, is served as an estimation of the uncertainty.

Our architecture is specifically designed for a $2\times$ upscaling factor in all three dimensions of the volumetric data (increase data size by a factor of $8\times$). With data augmentation and cross-scale training, as discussed in~\cref{sect:block_size_cross_scale}, we allow flexible super-resolution by recursively applying our trained model with a $2\times$ upscaling factor for each iteration.

\vspace{-9pt} 
\begin{algorithm}
\caption{Uncertainty Aware Super-resolution}\label{alg:uq}
    \SetKwInOut{KwIn}{Input}
    \SetKwInOut{KwOut}{Output}
    \KwIn{A low-resolution data $\mathbf{y}$
    }
    \KwOut{A super-resolved data $\bar{\mathbf{x}}$ and uncertainty estimation $\mathbf{x}_{\sigma}$}
    $\mathbf{z}_{lat}=e(\mathbf{y})$ \\
    $\mathbf{z}_{l}=\mathbf{z}_{lat}$ \\
    ${\mu}=m_{\mu}(\mathbf{z}_{l}) \ \textrm{,} \ {\Sigma}=m_{\Sigma}(\mathbf{z}_{l})$ \\
    \tcc{Sample Gaussian latent space $\mathbf{z}_0$ $N$ times, and estimate mean $\bar{\mathbf{x}}$ and variance $\mathbf{x}_{\sigma}$ of high-resolution outputs.}
    \For{$i \leftarrow 1$ \KwTo $N$}{
        $\mathbf{z}_0^{(i)}\sim \mathcal{N}(\mu, \Sigma)$ \\
        $\mathbf{z}_{h}^{(i)}=h^{-1}(\mathbf{z}_0^{(i)}, \mathbf{z}_{l} )$ \\
        $\mathbf{x}^{\prime(i)}=g^{-1}([\mathbf{z}_{h}^{(i)},\mathbf{z}_{l}])$
        \tcp{$\mathbf{x}^{\prime(i)}$ is a high-res output}
        $X.append(\mathbf{x}^{\prime(i)})$
    }
    $\bar{\mathbf{x}}=\frac{1}{N}\sum_{i=1}^N X(i)$ \\ 
    $\mathbf{x}_{\sigma} = \sqrt{\frac{1}{N} \sum_{i=1}^N (X(i) - \bar{\mathbf{x}})^2}$\\
    \KwRet{$\bar{\mathbf{x}}$, $\mathbf{x}_{\sigma}$}
\end{algorithm}

\setlength\textfloatsep{0.1\baselineskip}
\vspace{-10pt} 

\subsubsection{Uncertainty Visualization}
To analyze uncertainties in the super-resolution results, we utilize three uncertainty visualization methods. 
The first method directly visualizes the estimated variation field ($\mathbf{x}_{\sigma}$). 
The second approach encodes output variations ($\mathbf{x}_{\sigma}$) on the isosurfaces using color. 
The third method is based on the Probabilistic Marching Cubes algorithm proposed by P{\"o}thkow et al.~\cite{pothkow2011probabilisticMC}. For each low-resolution input, $N$ possible high-resolution outputs are generated using~\cref{alg:uq}. The mean and covariance matrix for each voxel of the generated high-resolution outputs are computed. The level-crossing probability is calculated by randomly sampling $n$ instances from the computed multivariate Gaussian distribution and applying the level-crossing criteria. The probability is computed as $m/n$ where $m$ is the number of samples that pass through the voxel~\cite{pothkow2011probabilisticMC}. 

\subsection{Loss Functions} \label{sect:lossFunction}
We formulate the super-resolution problem as the conditional generation of high-resolution data based on low-resolution data utilizing a normalizing flow model $f_\theta$ parameterize by $\theta$. Note that $f_\theta$ is the combination of flow $g$ and $h$ in our approach. The optimal parameter $\theta$ can be found in a data-driven manner by feeding the network with $M$ high and low-resolution training data pairs $\{(\mathbf{x}^{(i)}, \mathbf{y}^{(i)})\}_{i=1}^M$.
By maximizing the log-likelihood of high-resolution $\mathbf{x}$ conditioned on low-resolution $\mathbf{y}$, i.e., $\log p(\mathbf{x} \mid \mathbf{y})$, the model learns the data generation process. 
So the first loss we want to minimize is the negative log-likelihood loss $\mathcal{L}_{sr}$:
\vspace{-3pt}
\begin{equation} \label{eq:logploss}
    \mathcal{L}_{sr} = \mathbb{E}_{\mathbf{x},\mathbf{y}}[-\log p(\mathbf{x}\mid \mathbf{y})]
\end{equation}
To ensure that one component of the normalizing flow's latent space (i.e., $\mathbf{z}_{l}$) encodes the low-resolution information, the second loss we want to minimize is the L1 loss between flow $g$'s latent representation $\mathbf{z}_{l}$ and low-resolution encoder $e$'s output $\mathbf{z}_{lat}$: 
\vspace{-3pt}
\begin{equation} \label{eq:latloss}
    \mathcal{L}_{lat} = \mathbb{E}_{\mathbf{x},\mathbf{y}}[\|\mathbf{z}_{lat}-\mathbf{z}_{l}\|_1]
\end{equation}
where $\mathbf{z}_{l}$ and $\mathbf{z}_{lat}$ can be computed from~\cref{eq:g(x)} and~\cref{eq:z_lat}, respectively. In our work, normalizing flow models ($g$ and $h$) and the encoder $e$ in {\sysname} are trained together  through a combination of these two losses. We formulate our total training loss as: 
\vspace{-3pt}
\begin{equation} \label{eq:loss}
    \mathcal{L} = \mathcal{L}_{sr} + \lambda\mathcal{L}_{lat}
\vspace{-3pt}
\end{equation}
Our optimization goal is to minimize loss $\mathcal{L}$. 
Since high-resolution reconstruction and low-resolution encoding have different learning difficulties, a hyperparameter $\lambda$ is used in our loss to balance their optimization. 
The first loss $\mathcal{L}_{sr}$ is used to optimize the parameters of flow $g$ and $h$ so that the data is successfully transformed into a predicted Gaussian latent space. 
Concurrently, the second loss $\mathcal{L}_{lat}$ is used to constrain $\mathbf{z}_{l}$ and optimize flow $g$ to minimize the difference between $\mathbf{z}_{l}$ and $\mathbf{z}_{lat}$. Meanwhile, $\mathcal{L}_{lat}$ will also optimize parameters of encoder $e$ to make $\mathbf{z}_{lat}$ more similar to $\mathbf{z}_{l}$. Although it may introduce a moving-target problem, these two losses in {\sysname} are not competing with each other like GANs. Instead, one loss constrains the optimization of the other so that the models are still consistently optimized to reduce both $\mathcal{L}_{sr}$ and $\mathcal{L}_{lat}$. 

\subsection{Implementation}
\subsubsection{Flow Architecture}\label{sect:flow_step}
The invertibility for probability computation of normalizing flow is assured by design. 
The invertible flow $g$ and $h$ are based on the widely used Glow model~\cite{kingma2018glow} which consists of several stacked flow steps. These flow steps are the key to the invertibility and the modeling capacity of the normalizing flow. 
By stacking multiple flow steps as shown in~\cref{fig:architecture}, the model has the ability to learn the transformations between the Gaussian latent space and the high-resolution data space. 


\subsubsection{Block-based Processing and Cross Scale Training} \label{sect:block_size_cross_scale}
We train and evaluate our model based on data blocks from multiple scales. During training, the raw data is downscaled by a factor of $2^{n}$ and $2^{n+1}$ ($n\ge 0$), and we sample pairs of high-resolution (with scale factor $2^{n}$) and the corresponding low-resolution blocks (with scale factor $2^{n+1}$) for training. 
\clr{The model is trained to learn upscaling the input of different resolutions by a factor of $2\times$.} 
This training strategy improves the model's ability and robustness to be applicable to a range of scales. \clr{During inference, since our model is entirely based on convolutional layers,} we can utilize one trained model recursively across multiple scales for flexible super-resolution, with each iteration upscaling the data by a scale factor of $2\times$. 

To reduce artifacts and errors at block boundaries, we extend each block with extra padding before feeding it into the network. We crop the reconstructed blocks and only keep the central regions for the final output. 
This also helps reduce the modeling complexity so that instead of learning and sampling from a high-dimensional Gaussian latent space for the entire domain, we focus on modeling the missing information in each local block using a smaller Gaussian latent space.


\section{Results} \label{sect:Results}
In this section, we evaluate {\sysname} for scientific data super-resolution both quantitatively and qualitatively, and compare it with three baseline methods. Moreover, we demonstrate {\sysname}'s superiority in terms of uncertainty quantification, a feature missing from the baseline methods. 

\vspace{-6pt}
\begin{table}[!ht]
\small
\caption{\clr{Dataset name, training data, validation data, the total number of data pairs for training, and high-resolution output block size.}}
\vspace{-8pt}
\centering
 \begin{tabular}{c|c|c|c|c} 
 Dataset & \clr{Train} & \clr{Validation} & \# Training & Block Size\\ [0.5ex] 
 \hline
 Vortex & \clr{ts05, ts10, ts12, ts17, ts20} & \clr{ts08} & 1500 & 16\\
 Nyx & \clr{id35, id66 and id88} & \clr{id20} & 1050 & 16 \\
 Combustion & \clr{ts077, ts080, ts121} & \clr{ts045} & 1110 & 30 \\
 Plume & \clr{ts427, ts432, ts436, ts437} & \clr{ts420} & 1600 & 16 \\
 \end{tabular}\label{table:dataset}
\vspace{-10pt}
\end{table}

\subsection{Dataset and Training Parameters}
We evaluated our method using four scientific datasets, listed in~\cref{table:dataset}.
\textbf{Vortex} is a simulation of vortex structure. The data resolution is 128$\times$128$\times$128 with 30 time steps. We randomly sampled 1500 pairs of high and low-resolution data blocks from 5 time steps of the vorticity magnitude field for training. 
\textbf{Nyx} is a cosmological simulation produced by Lawrence Berkeley National Laboratory. The log density field with a resolution 256$\times$256$\times$256 was used for evaluation. We randomly sampled 1050 pairs of blocks from 3 ensemble members for training. 
\textbf{Turbulent Combustion} is a combustion simulation produced by Sandia National Laboratories. It is a time-varying multivariate dataset with resolution 480$\times$720$\times$120 across 122 time steps. We use the vorticity magnitude field for experiments, where 1110 pairs of data blocks from 3 time steps are used for training. 
\textbf{Plume} is a solar plume simulation with resolution 128$\times$128$\times$512. We utilize the velocity magnitude field for experiments. 1600 pairs of high and low-resolution blocks from 4 time steps are used for training. 
For detailed information about the specific time steps or ensemble members utilized for training and validation, please refer to~\cref{table:dataset}.

\clr{To demonstrate the generalization ability of {\sysname}, we conduct tests on all time steps that were unseen during training and validation for time-varying data and all unseen ensemble members for the ensemble simulation.  
To compare {\sysname}'s performance with the baseline methods and to provide a concise summary of the testing results, we select a representative test time step and an ensemble member to report and visualize the test results.}
\clr{See supplemental material Section 1 for more details.} 
Note that instead of a single scale, training of all datasets uses blocks from multiple scales.
\clr{
Although our method experiments with scalar datasets, it has the potential to be extended to bivariate, multivariate, or vector datasets by increasing the model's channel size.}

Our {\sysname} model is implemented using PyTorch$\footnote{https://pytorch.org}$ and trained on a single NVIDIA Tesla A100 GPU. The Adam optimizer~\cite{Diederik2015Adam} is used for all datasets, and the learning rate is $10^{-4}$ for all models. 

\subsection{Baselines}
We compare {\sysname} with three baseline methods: \clr{trilinear interpolation}, ESRGAN~\cite{wang2018esrgan}, and SSR-TVD~\cite{Han2022SSR-TVD}. \clr{Trilinear interpolation}, while simplistic, is a frequently used method to scale up data resolution. SSR-TVD and ESRGAN are all state-of-the-art deep learning based super-resolution methods. 
SSR-TVD is proposed for scientific data super-resolution and ESRGAN is widely adopted for image super-resolution. 
We found SSR-TVD performs better when trained without the discriminator. Therefore, we use the original SSR-TVD model as well as a modified version without discriminator, denoted as SSR-TVD (w/o D), as our baselines. For ESRGAN, to keep a similar model size to {\sysname} without affecting its performance, we use 10 stacked residual in residual dense blocks (RRDB)~\cite{wang2018esrgan} as our baseline. 
To ensure a fair cross-scale comparison, we modify SSR-TVD (w/ and w/o D) and ESRGAN to perform $2\times$ (in each of the three data dimensions, so a total of $8\times$) upscaling in one forward pass of the model. The output channel size of the last VoxelShuffle~\cite{shi2016PixelShuffle} layer in the generator of SSR-TVD is changed and only one VoxelShuffle layer is used in ESRGAN for upsampling. All other components of SSR-TVD are kept the same as the original implementation. The model sizes are shown in~\cref{tab:model_size}. 

We train SSR-TVD, SSR-TVD (w/o D), and ESRGAN similarly as~{\sysname} with the same training data blocks across different scales. These models are constructed to learn the relationship between high and low-resolution blocks regardless of their scale levels. 
\vspace{-5pt}
\begin{table}[!ht]
\small
\caption{Model size (G: generator, D: discriminator), total training time, and testing time for an upscale factor of $2\times$ for Vortex data.}
\vspace{-8pt}
\centering
 \begin{tabular}{c c c c}
 Method & Size & Train Time & Test Time\\ [0.5ex] 
  \hline
  \makecell{SSR-TVD} & \makecell{G: 51.4 MB \\ D: 10.2 MB} & \makecell{23h26m} & \makecell{12s} \\
  \cline{1-4}
  \makecell{SSR-TVD (w/o D)} & \makecell{G: 51.4 MB} & \makecell{23h4m} & \makecell{12s} \\
  \cline{1-4}
  \makecell{ESRGAN} & \makecell{G: 20.9 MB \\ D: 1.79 MB} & \makecell{23h43m} & \makecell{5s} \\
  \cline{1-4}
  \makecell{{\sysname} (our)} & \makecell{NF: 15.7 MB\\Encoder: 902 KB} & \makecell{23h48m} & \makecell{4s} \\
  \cline{1-4}
 \end{tabular}
 \label{tab:model_size}
\vspace{-14pt}
\end{table}

\subsection{Quantitative Evaluation} \label{sect:QuantiEval}
In this section, we quantitatively compare {\sysname}’s super-resolution results with three baseline methods, i.e., trilinear interpolation, SSR-TVD (w/ and w/o D), and ESRGAN. 

\begin{table*}[!ht]
\small
\vspace{-4pt}
\caption{PSNR and volume rendering image SSIM for {\sysname} and baselines' SR output. The best ones within 2\% difference are highlighted in bold.}
\vspace{-6pt}
\centering
 \begin{tabular}{c l |c c|c c|c c} 
 Data & Method & $\uparrow$PSNR ($2\times$) & $\uparrow$SSIM ($2\times$) & $\uparrow$PSNR ($4\times$) & $\uparrow$SSIM ($4\times$) & $\uparrow$PSNR ($8\times$) & $\uparrow$SSIM ($8\times$) \\ [0.5ex] 
  \hline
  \multirow{5}*{Vortex} & Lerp & 37.0220 & 0.9881 & 27.4537 & 0.9121 & 21.6235 & 0.8218 \\ 
  ~ & SSR-TVD & 19.8935 & 0.8033 & 16.7406 & 0.7487 & 15.4956 & 0.7336 \\ 
  ~ & SSR-TVD(w/o D) & \textbf{47.0051} & \textbf{0.9982} & \textbf{35.9391} & \textbf{0.9759} & \textbf{25.1956} & \textbf{0.8803} \\ 
  ~ & ESRGAN & \textbf{47.4034} & \textbf{0.9988} & \textbf{35.6226} & \textbf{0.9755} & \textbf{24.9267} & \textbf{0.8804} \\
  ~ & {\sysname} (our) & \textbf{47.2484} & \textbf{0.9980} & \textbf{35.5412} & \textbf{0.9726} & 24.4810 & \textbf{0.8751} \\
  \cline{1-8}

  \multirow{5}*{Nyx} & Lerp & 35.5707 & 0.9428 & 28.5468 & 0.8224 & 25.2392 & 0.7327 \\ 
  ~ & SSR-TVD & 22.2846 & 0.7015 & 20.5514 & 0.7495 & 19.5836 & 0.7325 \\ 
  ~ & SSR-TVD(w/o D) & 38.2107 & \textbf{0.9835} & \textbf{30.9282} & \textbf{0.8984} & \textbf{26.2338} & \textbf{0.7521} \\ 
  ~ & ESRGAN & 37.1247 & \textbf{0.9687} & 30.2658 & \textbf{0.8861} & 25.9154 & \textbf{0.7634} \\
  ~ & {\sysname} (our) & \textbf{39.6287} & \textbf{0.9669} & \textbf{30.4229} & 0.8616 & \textbf{26.0397} & \textbf{0.7519} \\
  \cline{1-8}

  \multirow{5}*{Combustion} & Lerp & 43.4335 & 0.9784 & 34.9869 & 0.8737 & 30.4519 & 0.7557 \\ 
  ~ & SSR-TVD & 20.8376 & 0.3943 & 17.7482 & 0.3096 & 16.0370 & 0.2944 \\ 
  ~ & SSR-TVD(w/o D) & 40.3971 & \textbf{0.9804} & 35.4757 & 0.8930 & 30.7108 & 0.7522 \\ 
  ~ & ESRGAN & \textbf{48.4751} & \textbf{0.9919} & \textbf{38.7662} & \textbf{0.9310} & \textbf{32.0014} & \textbf{0.7825} \\
  ~ & {\sysname} (our) & \textbf{48.5449} & \textbf{0.9927} & 37.7999 & \textbf{0.9232} & \textbf{31.5086} & \textbf{0.7892} \\
  \cline{1-8}
  \multirow{5}*{Plume} & Lerp & \textbf{45.9200} & \textbf{0.9957} & \textbf{38.7633} & \textbf{0.9836} & 33.3626 & 0.9596 \\ 
  ~ & SSR-TVD & 35.1342 & 0.9348 & 30.7222 & 0.9092 & 27.6458 & 0.8929 \\ 
  ~ & SSR-TVD(w/o D) & 44.3590 & 0.9853 & 36.6536 & 0.9303 & 34.0981 & 0.9153 \\ 
  ~ & ESRGAN & \textbf{45.3892} & \textbf{0.9924} & \textbf{38.8778} & \textbf{0.9796} & 34.1121 & \textbf{0.9608} \\
  ~ & {\sysname} (our) & \textbf{45.0816} & \textbf{0.9952} & \textbf{39.2172} & \textbf{0.9843} & \textbf{35.1595} & \textbf{0.9700} \\
   \cline{1-8}
 \end{tabular}
 \label{tab:eval}
\vspace{-13pt}
\end{table*}

\subsubsection{Evaluation Metrics}
For data level evaluation, we adopt peak signal-to-noise ratio (PSNR) to measure the difference between generated high-resolution data and the ground truth. For image level evaluation, we use the structural similarity index measure (SSIM)~\cite{wang2004ssim} to evaluate the quality of  volume rendering images. Higher PSNR and SSIM means better quality. 

\subsubsection{Evaluation on Different Upscaling Factors}
Instead of a static upscale factor, due to the cross-scale training strategy, we can upscale the low-resolution data multiple times with a $2\times$ scale factor for each iteration. In our experiments, we use upscale factors from $2\times$, $4\times$, to $8\times$. Note that hereafter the upscale factor means the increase in size per dimension, so the actual low-resolution volumetric data size is increased by factors of $8\times$, $64\times$, and $512\times$, respectively. 

In~\cref{tab:eval}, we evaluate the super-resolution results quantitatively in both data level (PSNR) and image level (SSIM) on four datasets. The results demonstrate that except for the SSR-TVD model, all other three deep learning based super-resolution methods outperform the trilinear interpolation for all upscale factors. The low performance of the SSR-TVD may be due to the instability during GAN's training, where the generator and the discriminator compete against each other. Specifically, through the loss plots we recorded during training, we found that the discriminator dominated the training so that it can always distinguish between the generated and the real data. As a result, the generator failed to improve its performance and cannot produce high-quality super-resolution output to fool the discriminator. When we removed the discriminator and only trained the generator, SSR-TVD (w/o D) performed much better compared to the original SSR-TVD model, as shown in the third row (i.e., SSR-TVD (w/o D)) of each dataset in~\cref{tab:eval}.  
The quality of the super-resolution results is largely affected by the GAN model's architecture, which poses a large concern for visualization scientists to trust GANs' results. In general, our {\sysname} has quantitatively comparable performance in both data level and image level for all upscale factors compared to SSR-TVD (w/o D) and ESRGAN for all datasets being evaluated. In~\cref{tab:eval}, the result also shows that for each test data, if we increase the upscale factor from $2\times$, $4\times$ to $8\times$, PSNR and SSIM will drop for all methods. This is due to the inevitable information loss in the low-resolution data, and thus the high-quality super-resolution becomes more challenging and uncertain. But the learning-based methods still demonstrate higher quality compared to the interpolation-based method. 

\begin{figure}[htp]
    \centering
    \includegraphics[width=0.8\columnwidth]{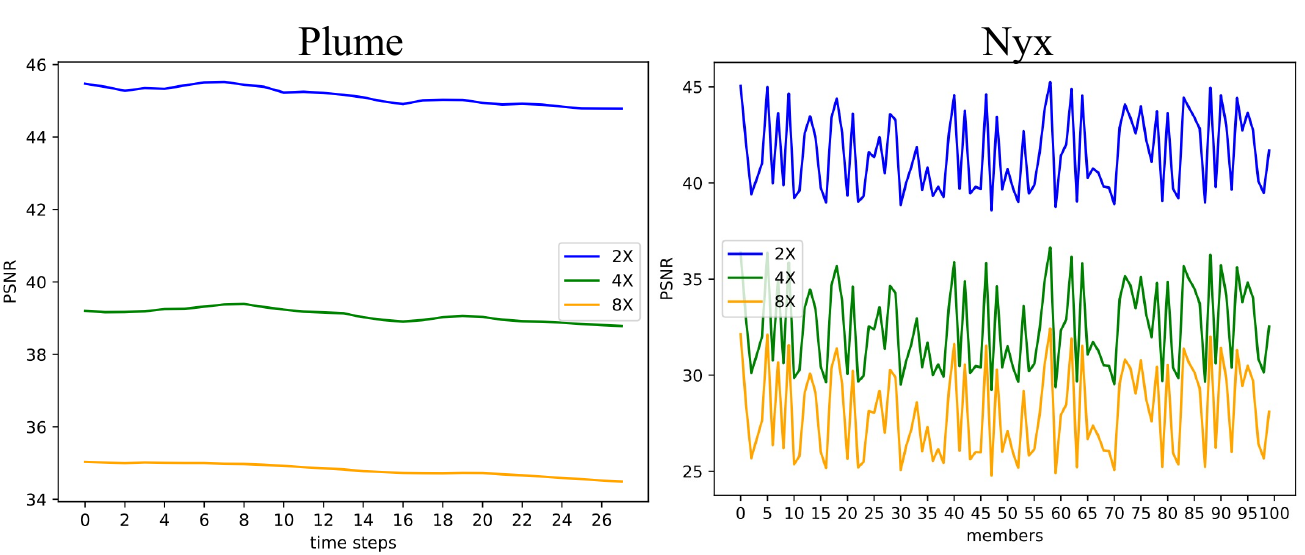}
    \vspace{-7pt}
    \caption{\clr{{\sysname}'s test PSNRs for all time steps of Plume data (left) and ensemble members of Nyx data (right).}}
    \label{fig:test_all_plume_nyx}
\vspace{-10pt}
\end{figure}

\clr{We conducted tests using {\sysname} on all time steps and ensemble members that were not used during training and validation. In~\cref{fig:test_all_plume_nyx}, the x-axis in each plot represents time steps or ensemble member IDs, while the y-axis corresponds to the PSNR values. 
The color encodes different upscaling factors, with blue, green, and yellow representing $2\times$, $4\times$, and $8\times$, respectively. The results show consistently high reconstruction quality with acceptable variations across all time steps for Plume data and all ensemble members for Nyx data.
These results demonstrate {\sysname}’s robust performance and its ability to generalize when presented with new low-resolution input. Due to the page limit, more testing results are in Section 2 of the supplemental material. } 

\vspace{-12pt}

We compare storage costs and training time of all learning-based super-resolution models, as shown in~\cref{tab:model_size}. The model size of SSR-TVD is $51.4$ MB for the generator and $10.2$ MB for the discriminator. For ESRGAN, it is $20.9$ MB for the generator and $1.79$ MB for the discriminator. While our proposed {\sysname} is the most lightweight model with a $15.7$ MB normalizing flow model and a $902$ KB encoder. Due to the smaller number of parameters, {\sysname} takes less time to train per epoch while a large model such as SSR-TVD is much slower. Besides, {\sysname} evaluates quicker than other large models.

\begin{figure}[t]
    \centering
    \includegraphics[width=\columnwidth]{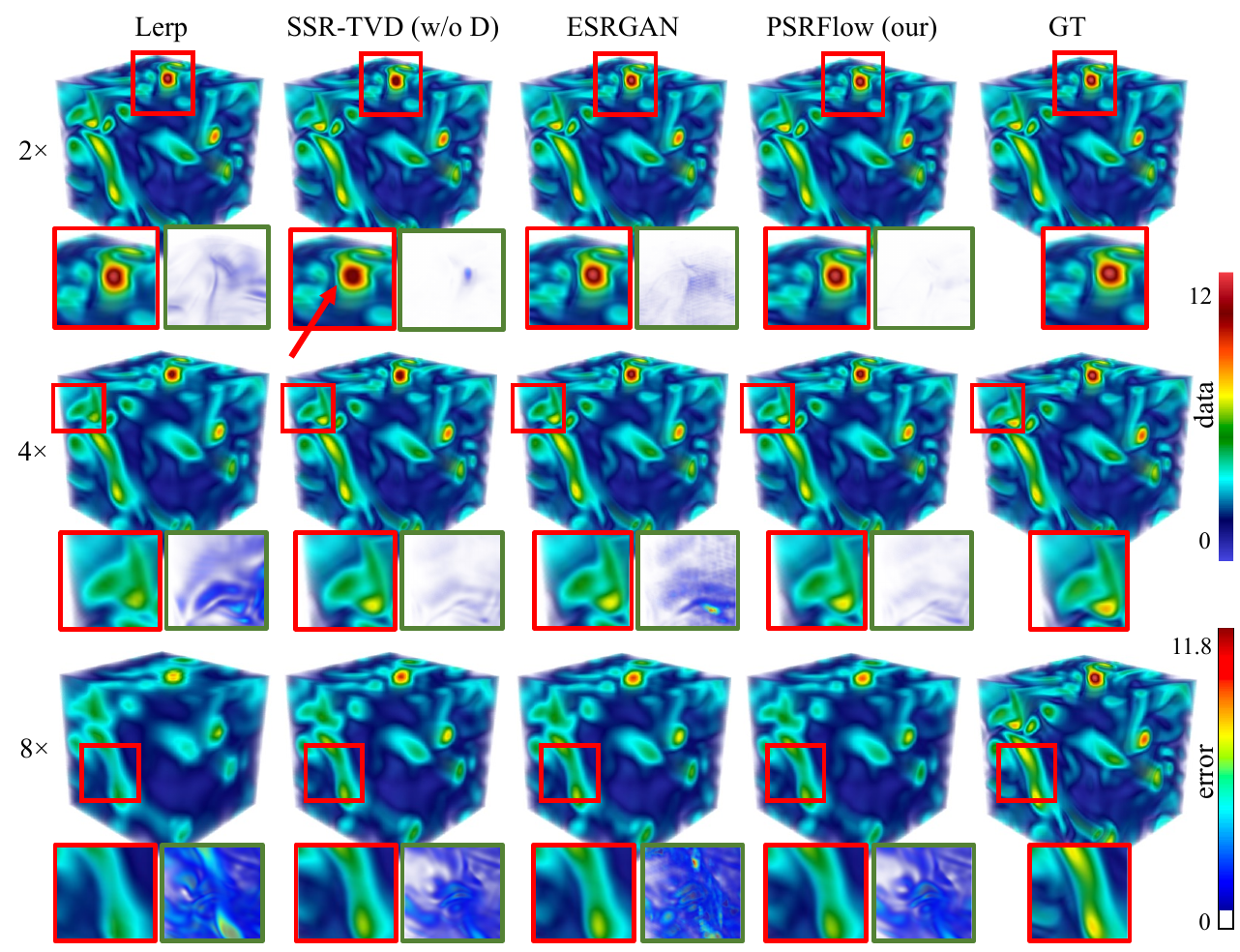}
    \vspace{-18pt}
    \caption{\clr{Volume rendering of baselines' and {\sysname}'s super-resolution results, error maps (green rectangle), and ground truth for Vortex data.}}
    \label{fig:vortex_vr}
\vspace{3pt}
\end{figure}

\begin{figure}[t]
    \centering
    \includegraphics[width=8.5cm]{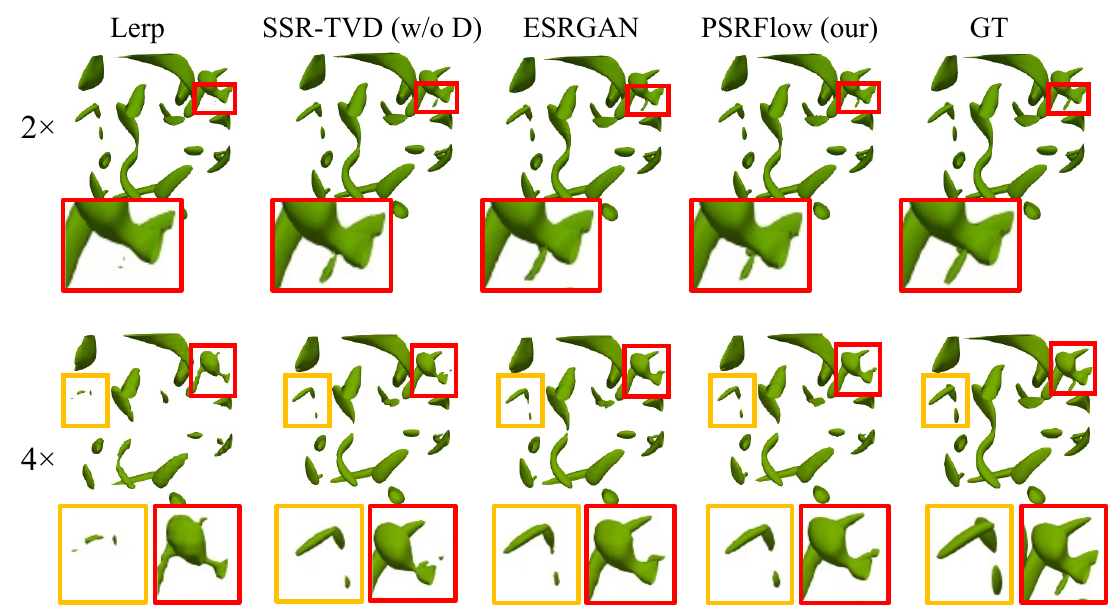}
    \vspace{-8pt}
    \caption{Isosurface rendering of baselines' and {\sysname}'s super-resolution results, and ground truth for Vortex data.}
    \label{fig:vortex_iso}
\vspace{-13pt}
\end{figure}

\begin{figure}[htp]
    \centering
    \includegraphics[width=\columnwidth]{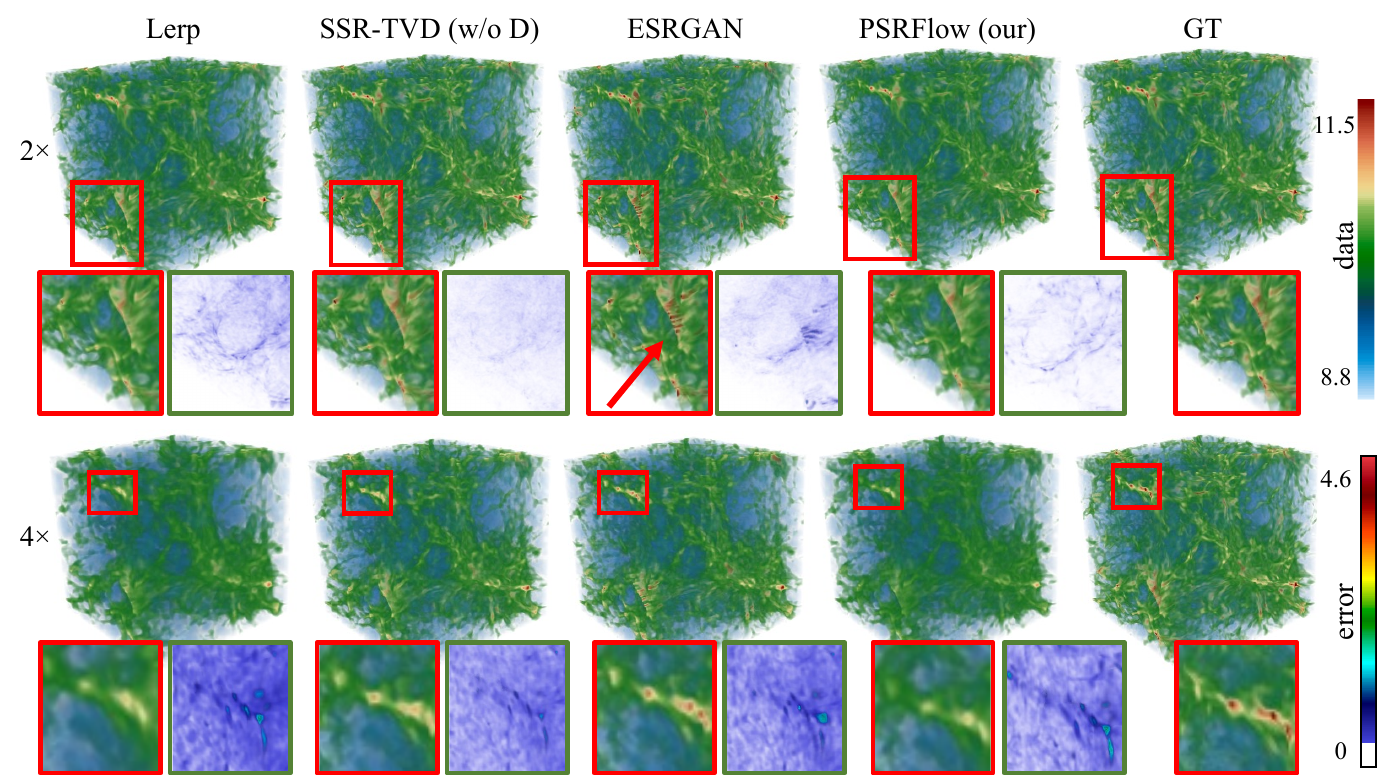}
    \vspace{-17pt}
    \caption{\clr{Volume rendering of baselines' and {\sysname}'s super-resolution results, error maps (green rectangle), and ground truth for Nyx data.}}
    \label{fig:Nyx}
\vspace{3pt}
\end{figure}

\subsection{Qualitative Evaluation}\label{sect:QualiEval}
We qualitatively compare the proposed {\sysname} with the baseline methods via volume rendering and isosurface rendering of super-resolved results. We use the same visualization setting for all rendering results of the same dataset. 

In~\cref{fig:vortex_vr}, we show volume rendering images of trilinear interpolation (Lerp), SSR-TVD (w/o D), ESRGAN, and {\sysname}'s super-resolution results for the Vortex dataset, and compare them with the high-resolution ground truth. The upscale factors from the top row to the bottom row are $2\times$, $4\times$, and $8\times$.
\clr{For each super-resolution result, we provide a zoom-in region (red rectangles) at the bottom of the volume rendering image, accompanied by the corresponding squared error map on its right side (green rectangles).} 
\clr{By comparing the volume rendering images for error maps of the proposed {\sysname} with the three baselines (the first three columns),} we found that the learning-based methods can produce super-resolved data with better high-frequency details, unlike the interpolation-based method which tends to produce blurry outputs.
Although SSR-TVD (w/o D), ESRGAN, and {\sysname} all produce high-quality results that are visually similar, SSR-TVD (w/o D) still contains artifacts, such as generating non-existing high vorticity magnitude voxels in the center of the vortex core, as shown in the zoom-in region of the first row ($2\times$) in the second column of~\cref{fig:vortex_vr}. However, the third row ($8\times$) shows relatively low quality of all methods, while interpolation has the worst performance. This is anticipated since at this level, most feature information is lost in the low-resolution data, making it difficult to reconstruct the original vortex core structures accurately.
We do not include SSR-TVD's qualitative evaluation here since it has low quality visually with obvious artifacts. Results of SSR-TVD are presented in the supplementary material.

\Cref{fig:vortex_iso} shows isosurface rendering images of trilinear interpolation, SSR-TVD (w/o D), ESRGAN, {\sysname}'s super-resolution results, and high-resolution ground truth for the Vortex data with $isovalue=6$. 
The upscale factors for the first and the second row are $2\times$ and $4\times$. 
In the predicted high-resolution isosurface rendering images, we observed that trilinear interpolation loses isosurface features easily such as the red highlighted region in the first row and the yellow highlighted region in the second row. This is likely due to the fact that the simple trilinear interpolation method can not reconstruct complex features accurately. 
In the red highlighted regions of the second row, we found SSR-TVD (w/o D) also has difficulties reconstructing some details and resulting in disconnected features. ESRGAN and {\sysname} have similar high-quality super-resolution results compared to the ground truth.

In \cref{fig:Nyx}, we compare volume rendering results of trilinear interpolation, SSR-TVD (w/o D), ESRGAN, and {\sysname}'s super-resolution results for the Nyx dataset. The upscale factors for the first and the second row are $2\times$ and $4\times$. It is clear that learning-based methods can produce improved details in super-resolved results than trilinear interpolation. Overall SSR-TVD (w/o D), ESRGAN, and {\sysname} produce similar high-quality results. However, ESRGAN can generate artifacts that do not exist in the original high-resolution data. For instance, in the red highlighted regions of the first row, ESRGAN (the third column, red arrow) produces non-existing high-density features. This poses a challenge for scientists to trust GAN's super-resolution outputs.

\begin{figure}[t]
    \centering
    \includegraphics[width=\columnwidth]{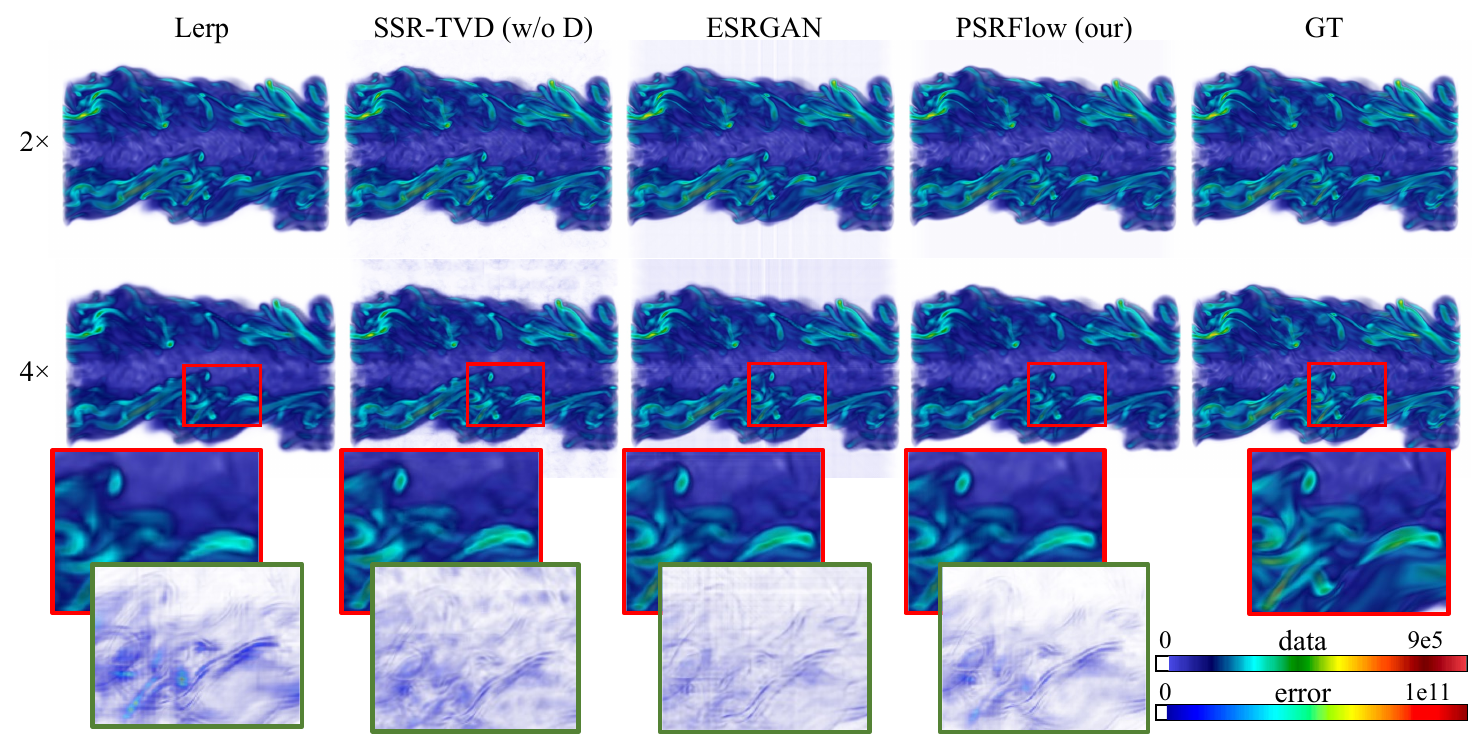}
    \vspace{-16pt}
    \caption{\clr{Volume rendering of baselines' and {\sysname}'s super-resolution results, error maps (green rectangle), and ground truth for Combustion.}}
    \label{fig:combustion}
    \vspace{3pt}
\end{figure}

\Cref{fig:combustion} shows volume rendering images of baselines' and {\sysname}'s super-resolution results, \clr{and error maps} for Combustion data. From the zoom-in regions in the second row, we can see that deep learning based methods can produce sharper reconstructions with higher-quality features compared to interpolation-based methods. However, \clr{when compared to {\sysname}, SSR-TVD (w/o D) and ESRGAN produce noticeable artifacts in both the error maps of zoom-in regions and the background area of the whole reconstruction's rendering images.}

In conclusion, SSR-TVD (w/o D), ESRGAN, and {\sysname} all can produce high-quality super-resolution results, however, since SSR-TVD (w/o D) and ESRGAN are all GAN-based models, they have training instability problems and are easy to generate artifacts. As a result, error analysis and uncertainty estimation have become crucial for scientists to gain trust to the super-resolution results. In this respect, {\sysname} outperforms all other methods with reliable uncertainty quantification. We focus on uncertainty quantification of {\sysname} in the next section. 

\vspace{-10pt}
\begin{figure}[htp]
    \centering
    \includegraphics[width=7cm]{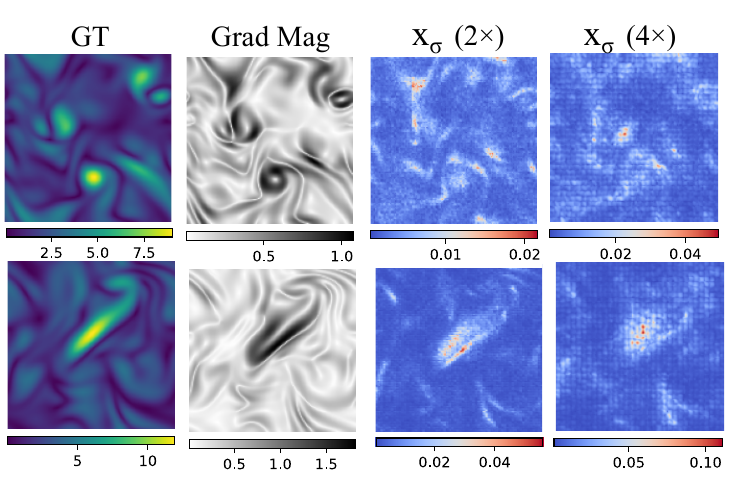}
    \vspace{-8pt}
    \caption{\clr{Uncertainty analysis of {\sysname}'s super-resolution outputs on Vortex data. Each row is a slice from the 3D data. The four columns are high-resolution ground truth (GT), its gradient magnitude (Grad Mag), and estimated uncertainties ($\mathbf{x}_{\sigma}$) for upscale factor $2\times$ and $4\times$. }}
    \label{fig:vortex_qu_v2}
\vspace{-17pt}
\end{figure}

\begin{figure}[htp]
    \centering
    \includegraphics[width=8.5cm]{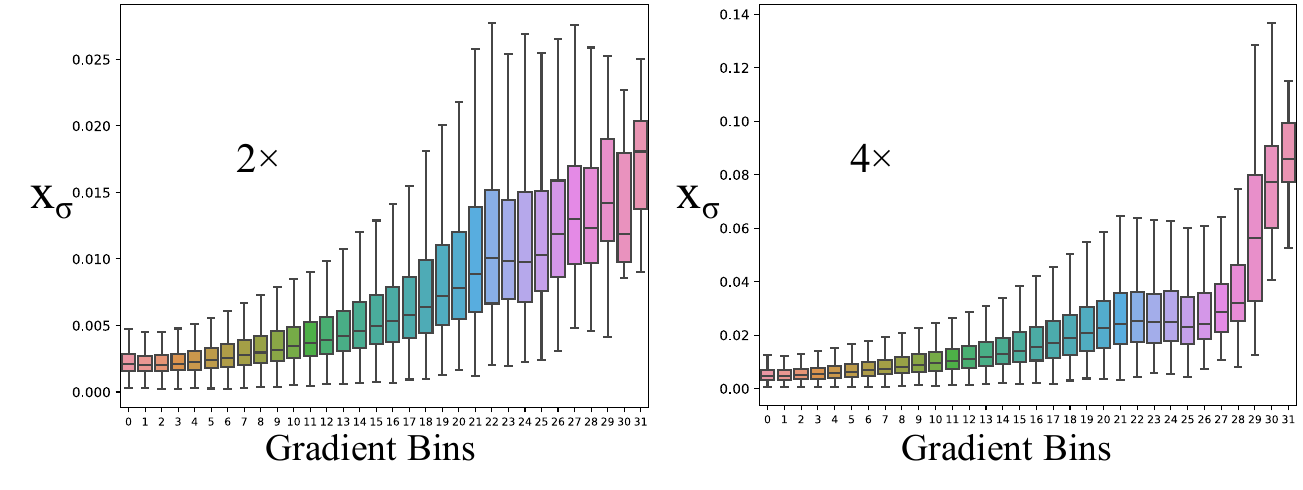}
    \vspace{-11pt}
    \caption{Uncertainty value distribution for each gradient magnitude bin. }
    \label{fig:vortexUQ_box_hex}
\vspace{-9pt}
\end{figure}


\begin{figure}[t]
    \centering
    \includegraphics[width=\columnwidth]{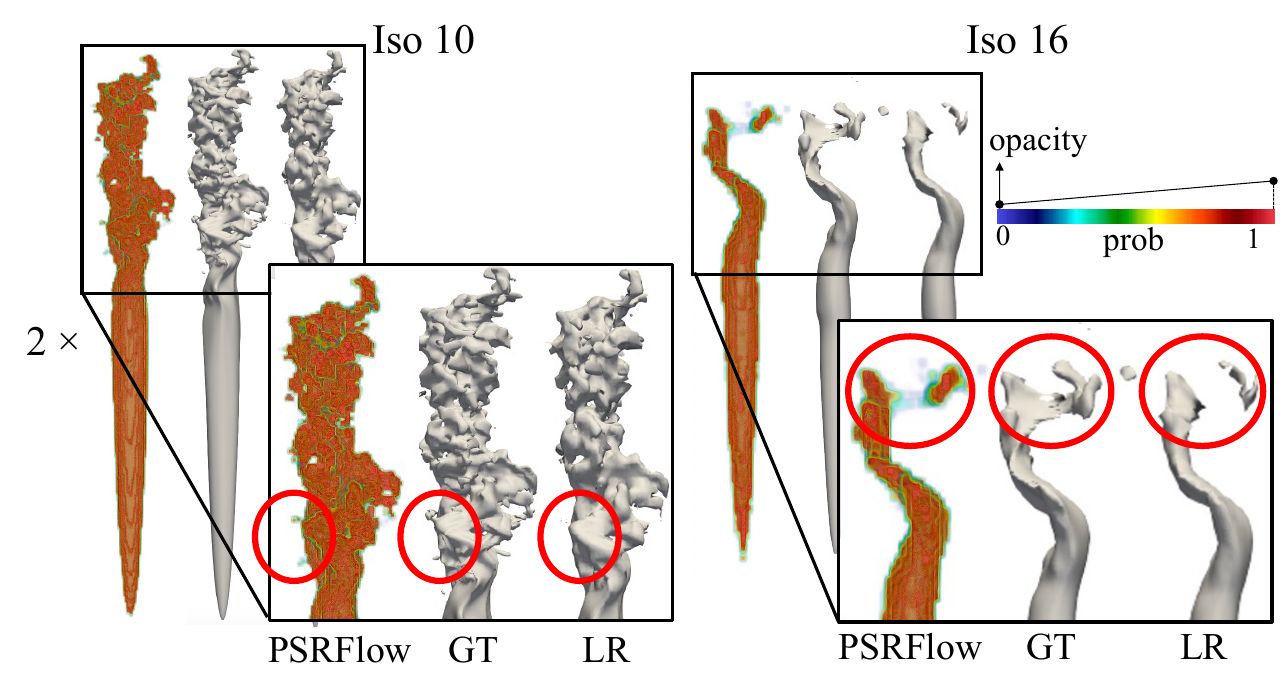}
    \vspace{-23pt}
    \caption{Level-crossing probability visualization and isosurface rendering of Plume data with $isovalue=10$ and $16$, with a scaling factor of $2\times$.}
    \label{fig:plume_pmc}
\vspace{3pt}
\end{figure}



\subsection{Uncertainty Quantification} \label{sect:UQ_result}
Due to the explicit distribution modeling, {\sysname} captures the variations of missing high-frequency information in the Gaussian latent space. Given a low-resolution input, {\sysname} allows exploring the corresponding super-resolved data space by sampling from the Gaussian latent space. Each sample will reconstruct one realization of the high-resolution data. By measuring and analyzing the variation among multiple realizations, we can estimate the uncertainties of {\sysname}'s super-resolution outputs. 
In this section, we estimate uncertainties via~\cref{alg:uq}, where we set the number of samples $N=40$. 

\subsubsection{Sources of Uncertainty}
In this section, we analyze the source of uncertainty. 

In~\cref{fig:vortex_qu_v2}, we show the results on the Vortex data. Each row in~\cref{fig:vortex_qu_v2} corresponds to a 2D slice from the 3D data. The first two columns are the high-resolution ground truth and its gradient magnitude field. 
\clr{The next two columns are the estimated uncertainties of the high-resolution data ($\mathbf{x}_{\sigma}$) for upscale factor $2\times$ and $4\times$, respectively.}
The gradient magnitude field uses a colormap from white to black whereas regions with higher gradient magnitude will be darker. \clr{$\mathbf{x}_{\sigma}$ use a colormap that goes from blue to red with white in the middle. }
In \cref{fig:vortex_qu_v2}, the variation in high-resolution data ($\mathbf{x}_{\sigma}$) reveals clear patterns of high-uncertainty regions. 
Comparing $\mathbf{x}_{\sigma}$ with the gradient magnitude field of the ground truth, we find that the high-uncertainty (high-variation) regions are likely to be regions with higher gradients. 
To verify this, we plot and analyze the correlation between the uncertainty values and the gradient magnitude values in~\cref{fig:vortexUQ_box_hex}. We compute the histogram of gradient magnitudes using $32$ bins. For each bin, we calculate the mean and variance of all voxels' uncertainties in this bin and plot them as a boxplot in~\cref{fig:vortexUQ_box_hex}. The x-axis of each boxplot represents the gradient magnitude bins, while the y-axis represents the uncertainty values. As the gradient magnitude increases, there is a clear trend that the mean uncertainty in each bin will also increase for both upscale factors $2\times$ (left) and $4\times$ (right). 

One possible reason for the correlation is that high-gradient regions tend to have more complex features than flat regions. When these regions are downsampled, the high-frequency gradient information gets smoothed out. Thus, one low-resolution input may correspond to multiple possible high-resolution outputs, and the variations occur at the complex regions (e.g., high-gradient regions). In this case, the model has lower confidence and higher uncertainties in complex regions of the super-resolution output. For simple regions, there are fewer variations and the model is more confident and less uncertain about the output. 

Comparing the spatial and value distributions of the variations in high-resolution data ($\mathbf{x}_{\sigma}$) for different upscale factors in \cref{fig:vortex_qu_v2} and \cref{fig:vortexUQ_box_hex}, we find that when the upscale factor is increased from $2\times$ to $4\times$, we have regions with increased uncertainties and the value range of uncertainties is also increased. 
\clr{This is probably because as the upscaling factor increases, the complexity of the super-resolution problem also increases, making it more challenging to estimate fine details due to the large information loss. There are more ambiguities during super-resolution, thus the level of uncertainty is increased. }

\clr{\cref{fig:plume_pmc} shows the level-crossing probabilities computed using the probabilistic marching cubes algorithm~\cite{pothkow2011probabilisticMC} for {\sysname}'s output, along with isosurface rendering images of the ground truth (GT) and the low-resolution data (LR) for the Plume dataset at $isovalue=10$ and $16$. 
The upscaling factor employed is $2\times$. We render the level-crossing probability field of {\sysname}'s super-resolution results, where red (opaque)  represents the high probability, and blue (transparent) represents the low probability. By comparing the probability field with the ground truth and low-resolution data, we observe that {\sysname} is able to reconstruct missing features in the low-resolution input. However, since the low-resolution information is incomplete, the model exhibits lower probabilities in these regions, as highlighted by the red circles, which indicate higher uncertainties. This is expected since the model is given limited input information and its performance heavily depends on the quality of the low-resolution input. }

In summary, the uncertainties come from the variations of possible high-resolution data. When the model is not sure about the exact super-resolved data it should produce based on the low-resolution input, it will have higher uncertainties in these high-variation regions.

\begin{figure}[t]
    \centering
    \includegraphics[width=5.5cm]{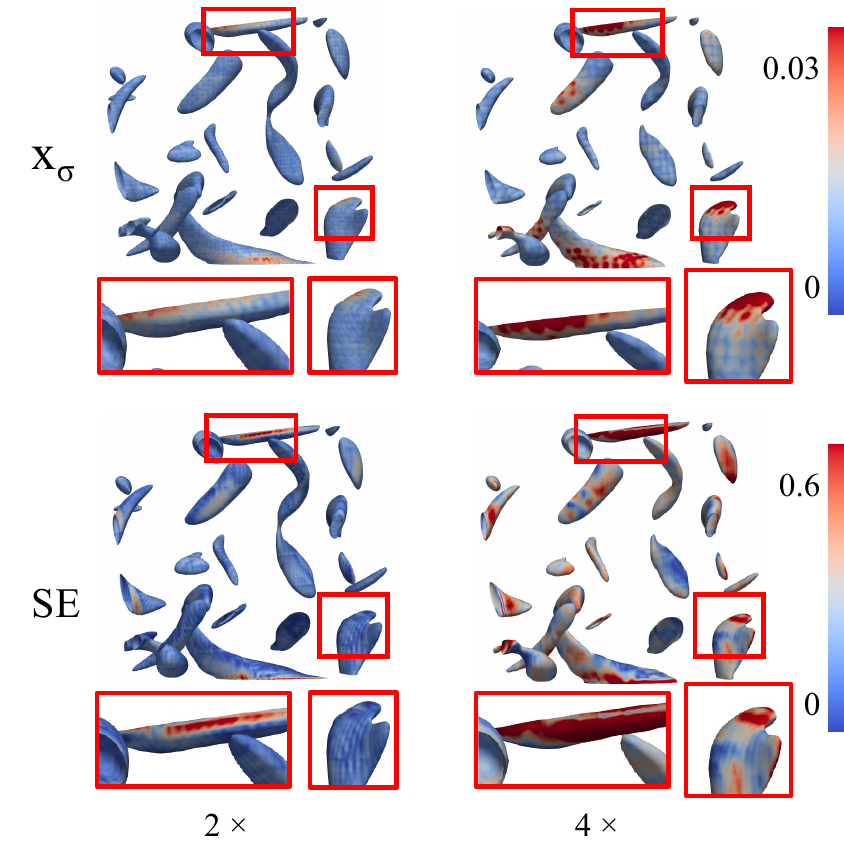}
    \vspace{-15pt}
    \caption{\clr{Isosurface uncertainty visualization (first row) and isosurface squared error visualization (second row) of Vortex data with $isovalue=6$.}}
    \label{fig:vortex_iso_std}
\vspace{-10pt}
\end{figure}

\begin{figure}[t]
    \centering
    \includegraphics[width=\columnwidth]{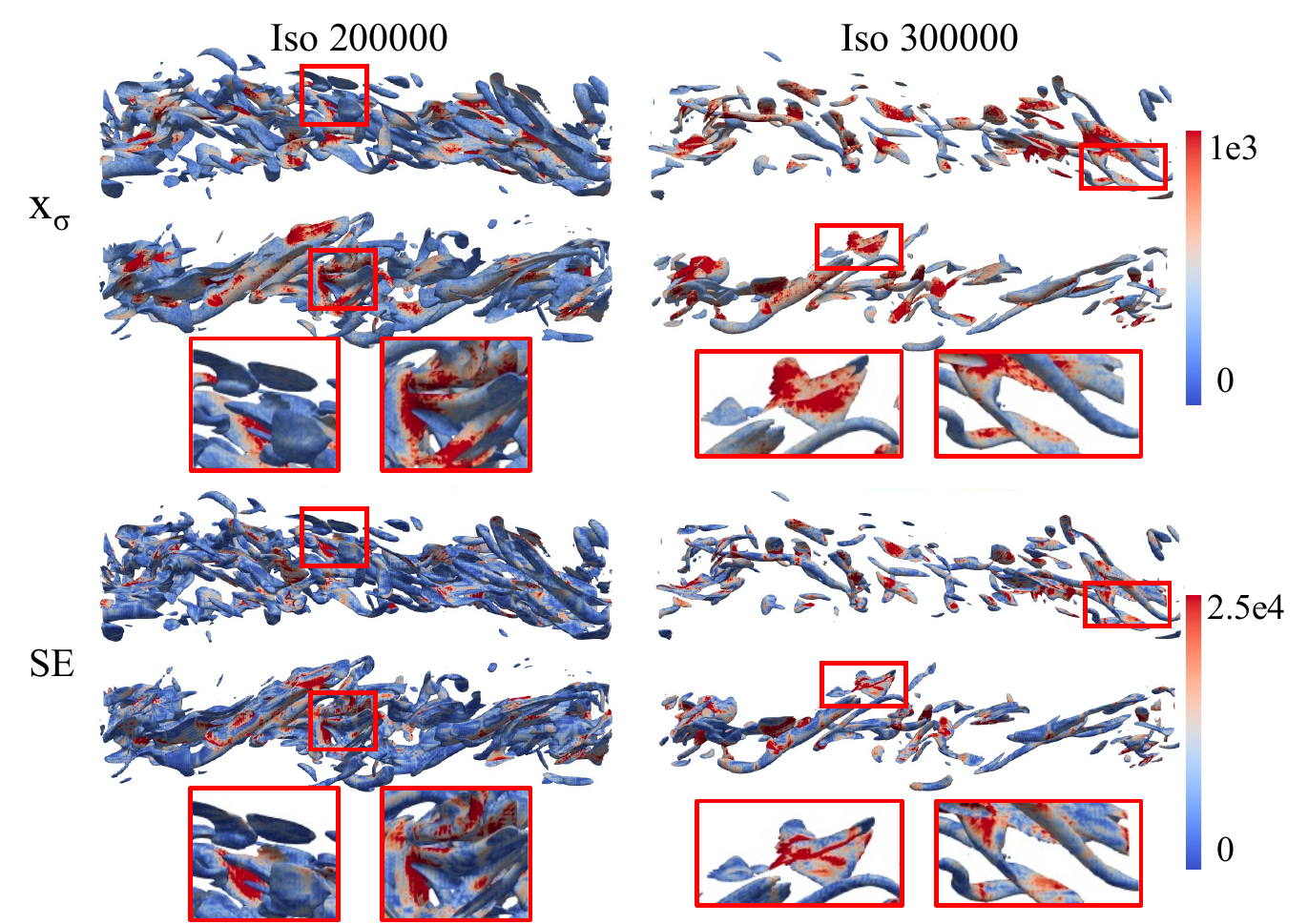}
    \vspace{-16pt}
    \caption{\clr{Isosurface uncertainty visualization (first row) and isosurface squared error visualization (second row) of the Combustion dataset. }}
    \label{fig:combustion_iso_std}
\vspace{4pt}
\end{figure}

\subsubsection{Uncertainty for Error Analysis}
In this section, we analyze the correlation between errors and uncertainties in {\sysname}'s outputs \clr{on the Vortex and Combustion data.}

\clr{\cref{fig:vortex_iso_std} shows isosurface uncertainty and isosurface error visualization for the Vortex dataset with $isovalue=6$. The error refers to the squared error calculated between the mean of the {\sysname}'s outputs and the high-resolution ground truth. Uncertainties ($\mathbf{x}_{\sigma}$) and squared errors (SE) are encoded directly on the surfaces using color, with red indicating higher values. The first row shows isosurfaces uncertainties and the second row shows squared error maps. The columns are results for scaling factors of $2\times$ and $4\times$, respectively. 
When comparing the uncertainties of $2\times$ and $4\times$ results, we also observe an increase in uncertainty in many regions as the upscaling factor increases. 
Meanwhile, our uncertainty shows blob-like artifacts, this is due to the sampling in the latent space. As the latent space provides a wider receptive field, instead of pixel-wise uncertainty, we have blob-like patterns in uncertainty.
By comparing the zoom-in regions in red rectangles across rows, we found the regions with high uncertainty also exhibit higher squared errors. }
This is because high-uncertainty regions are areas with high ambiguity and high variation due to the information loss, the model is less confident in these regions and errors are more likely to occur in these highly uncertain regions. This demonstrates that we can utilize uncertainty as an indication of the quality of super-resolution outputs. 


\clr{\cref{fig:combustion_iso_std} shows isosurface uncertainty and isosurface error visualization for the Combustion dataset. The two columns represent results for $isovalue=200000$ and $300000$, respectively. Similar to previous findings, the red rectangles in the rendering images also reveal a correlation between uncertainty ($\mathbf{x}_{\sigma}$) and error (SE) for each isovalue. }

During post-hoc analysis, when the ground truth high-resolution data is not available, scientists cannot assess the quality of super-resolution outputs. 
Reliable uncertainty quantification is crucial in this case since it provides scientists with indications about what degree of confidence the scientists should have toward the super-resolution results. \clr{In the future, to improve the model's performance, one can augment the training data in these highly uncertain regions to reduce testing errors. }
Previous learning-based super-resolution methods, despite their high performance, lack the ability for error estimation and uncertainty quantification, which can be offered by the proposed {\sysname}.


\section{Discussion and Future work} 
Normalizing flow is attractive due to the effective probabilistic distribution modeling and uncertainty quantification. We have demonstrated {\sysname} can be used for spatial super-resolution with uncertainty quantification. There are still several limitations to our work. 

First, the super-resolution quality is largely affected by the quality of the low-resolution data which can be improved by considering data importance during downsampling. 
Also, we do not guarantee temporal coherence for the super-resolved outputs. In the future, we would like to extend our approach to the temporal domain and quantify the uncertainties caused by the temporal super-resolution of features. 

Second, normalizing flows have limitations such as network capacity and dimensionality constraints. So instead of invertible modeling between a complex data space and a Gaussian latent space, we can reduce the data size by encoding data into a latent space and using the flow model to learn the transformation between a complex latent space and a Gaussian latent space for efficient data generation and likelihood estimation. This also reduces the computational cost of training deep and complex normalizing flows.


Third, our goal is to add high-frequency details to the low-resolution data through stacked normalizing flow steps. However, this is still an estimation imposed by the training loss. One direction to improve is to restore the missing high-frequencies explicitly in the Fourier domain. For example, utilize normalizing flows based on wavelets. 

Uncertainty quantification with normalizing flows is a promising research direction for deep-learning-based scientific visualization. In the future, we would like to use flow models to improve the reliability of neural network outputs for various scientific visualization applications. 



\section{Conclusion} 
In this paper, we propose {\sysname}, a novel deep learning based super-resolution algorithm with uncertainty quantification. Our work is based on normalizing flows to model the relationships between low and high-resolution data. The missing high-frequency information is captured in the Gaussian latent space. The Gaussian distribution allows efficient exploration of the high-resolution space given the low-resolution input. By sampling from the Gaussian latent space multiple times, one can compute the variations of the high-resolution data for uncertainty estimation. The quantified uncertainties can serve as indications for potential errors in the super-resolved results when the ground truth is no longer available. With cross-scale training, we can utilize one trained model recursively for flexible super-resolution of different scales. Our experimental results demonstrate that our model can achieve high-quality super-resolution and effective uncertainty quantification. 

\newpage
\section*{Supplemental Materials}
\label{sec:supplemental_materials}
Our supplemental material is available on OSF at \href{https://osf.io/b3j5d/?view_only=4fcd9294bded4d11ae6980623d29c708}{anonymous link}, which includes training/testing/validation dataset settings, additional testing results, {\sysname}'s results on noisy input, a table of model size and training time, a table of concepts and notations, description of flow architecture, hyperparameter analysis, and volume rendering images of original low-resolution inputs and their super-resolution outputs using baselines and PSRFlow for all datasets with scale factors $2\times$, $4\times$, and $8\times$.

\section*{Figure Credits}
\label{sec:figure_credits}
\Cref{fig:bg_nf} image credit: a modified version of normalizing flow produced by Riebesell Janosh, "Random TikZ Collection", 2022, GitHub repository, licensed under the MIT license. Retrieved from https://github.com/janosh/tikz.


\acknowledgments{
This work is supported in part by the US Department of Energy SciDAC program DE-SC0021360 and DE-SC0023193, National Science Foundation Division of Information and Intelligent Systems IIS-1955764, and Los Alamos National Laboratory Contract C3435.
}

\bibliographystyle{abbrv-doi-hyperref}

\bibliography{template}

\appendix 

\end{document}


\maketitle

\section{\clr{Experiment Dataset Settings}}
\clr{
To build a reliable model, we split our dataset for training, validation, and testing. Given a time-varying dataset, we train the model based on randomly sampled blocks from several selected training time steps. For the ensemble simulation dataset, the training data are from several selected ensemble members. One time step or ensemble member that was unseen during training was randomly chosen for validation. \Cref{table:train_val} provides detailed information about the datasets used for training and validation. For example, for the Vortex dataset, which is a time-varying dataset with 30 time steps, we randomly sampled 1500 pairs of high and low-resolution data blocks from 5 time steps (5, 10, 12, 17, and 20) for training. Blocks from a randomly selected time step (8) were used for validation. }

\clr{
To demonstrate the generalization ability of {\sysname}, we conduct tests on all time steps that were not used during training and validation for time-varying datasets, and on all unseen ensemble members for the ensemble simulation dataset.  
To compare {\sysname}'s performance with the baseline methods, we report the test results on one testing time step for the time-varying data and one ensemble member for the ensemble simulation in the main text. We only show the test results for one time step or one ensemble member in the quantitative and qualitative evaluation of {\sysname} in the main text mainly because we aim to provide a concise summary of the representative testing results. 
The selected time step and ensemble member server as representative examples across all testing data. More specifically, \cref{table:test_compare} provides detailed information about the datasets used for comparison with baseline methods.}

\begin{table}[!ht]
\small
\caption{\clr{Dataset name, training data, validation data, the total number of data pairs for training, and high-resolution output block size.}}
\centering
 \begin{tabular}{c|c|c|c|c} 
 Dataset & Train & Validation & \# Training & Block Size \\ [0.5ex] 
 \hline
 Vortex & ts05, ts10, ts12, ts17, ts20 & ts08 & 1500 & 16 \\
 Nyx & id35, id66 and id88 & id20 & 1050 & 16 \\
 Combustion & ts077, ts080, ts121 & ts045 & 1110 & 30 \\
 Plume & ts427, ts432, ts436, ts437 & ts420 & 1600 & 16 \\
 \end{tabular}
 \label{table:train_val}
\end{table}

\vspace{-2pt}

\begin{table}[!ht]
\small
\caption{\clr{The data selected for comparison with baselines.}}
\centering
 \begin{tabular}{c|c} 
 Dataset & Compare with baselines \\ [0.5ex] 
 \hline
 Vortex & ts01 \\
 Nyx & id4 \\ 
 Combustion & ts053 \\
 Plume & ts441  \\ 
\end{tabular}\label{table:test_compare}
\vspace{-4pt}
\end{table}

\section{\clr{More Testing Results}}
\clr{
In this section, we report {\sysname}'s test results on all time steps that were not used during training and validation for the time-varying datasets (i.e., Vortex, Combustion, and Plume datasets).  
To evaluate how {\sysname} performs when the simulation parameter changes, we show the test results on all ensemble members with different simulation parameters that were not used during training and validation for the ensemble simulation dataset (Nyx dataset).
During testing, {\sysname} takes the low-resolution data as input and outputs high-resolution reconstruction. We utilize PSNR to measure the quality of the generated high-resolution data. }

\clr{
In \cref{table:test_psnr}, we present the mean and standard deviation of PSNRs for all testing time steps for the time-varying dataset and all ensemble members for the ensemble simulation. 
\Cref{fig:test_psnr} displays line plots of all testing PSNRs for all datasets. In each plot, the x-axis represents time steps or ensemble member IDs, while the y-axis corresponds to the PSNR values. The color encodes different upscaling factors, with blue, green, and yellow representing $2\times$, $4\times$, and $8\times$, respectively.
The results show consistently high reconstruction quality with acceptable variations across all time steps of Vortex, Plume, and Combustion datasets, and all ensemble members of the Nyx dataset. These results demonstrate the model’s robust performance and its ability to generalize when presented with new low-resolution input. }

\clr{
In the results for the Combustion data on the bottom-right of \cref{fig:test_psnr}, we observe a sudden drop in PSNR. We find this is due to the occurrence of the combustion phenomenon, characterized by an abrupt increase in vorticity magnitude that was not learned during training. This negatively impacts the quality of a specific local region and reduces the overall PSNR score of the super-resolution results. Despite the drop, the PSNR remains at a relatively acceptable level. Furthermore, the performance can be enhanced by including additional training samples that cover the extreme data distribution present in these time steps. 
}

\begin{table}[t]
\small
\caption{\clr{{\sysname}'s testing PSNR mean and standard deviation of all time steps (or ensemble members).}}
\centering
 \begin{tabular}{c|c|c|c} 
 Dataset & scale factor & PSNR (mean) & PSNR (std) \\ [0.5ex] 
 \hline
 \multirow{3}*{Vortex} & $2\times$ & 47.0989 & 0.2955 \\
 ~ & $4\times$ & 35.3517 & 0.2107\\
 ~ & $8\times$ & 24.3137 & 0.1824\\
 \hline
 \multirow{3}*{Nyx} & $2\times$ & 41.5621 & 2.0228 \\
 ~ & $4\times$ & 32.4884 & 2.2375\\
 ~ & $8\times$ & 28.1014 & 2.2954\\
 \hline
 \multirow{3}*{Combustion} & $2\times$ & 46.8265 &  7.4598 \\
 ~ & $4\times$ & 39.0197 & 3.7553\\
 ~ & $8\times$ & 32.5738 & 2.6106\\
 \hline
 \multirow{3}*{Plume} & $2\times$ & 45.1378 &  0.2397 \\
 ~ & $4\times$ & 39.0861 & 0.1744\\
 ~ & $8\times$ & 34.8144 & 0.1679\\
\end{tabular}\label{table:test_psnr}
\vspace{-10pt}
\end{table}

\begin{figure}[htp]
    \centering
    \includegraphics[width=\columnwidth]{figures/test_all.pdf}
    \vspace{-12pt}
    \caption{\clr{{\sysname}'s testing PSNRs for all time steps of Vortex, Plume, and Combustion datasets, and for all ensemble members of the Nyx dataset.}}
\vspace{-5pt}
\label{fig:test_psnr}
\end{figure}

\section{\clr{Noisy Input}}
\clr{
To test {\sysname}'s robustness on noisy input, we conduct additional experiments with varying levels of noise injected into the low-resolution input during testing. The injected noises are sampled from a Gaussian distribution with zero mean. The standard deviation of this Gaussian distribution is determined by scaling the noise percentage with the data range. The resulting noise is then added to the original low-resolution data. 
Noise levels vary from $1\%$ to $10\%$ for Vortex, Plume, and Combustion datasets, while for the Nyx dataset, the range is set to $1\%$ to $5\%$ due to a significant performance drop beyond that point.}

\clr{
We evaluate and report {\sysname}'s testing performance in reconstructing the high-resolution data for each dataset. As shown in~\cref{fig:nosiy_input}, the x-axis in each plot represents noise levels and the y-axis represents the corresponding output PSNR values. 
Our result shows that the performance is influenced by the noise level, and as the noise level increase, the model's performance drops for each dataset. However, although being trained on clean data, our model exhibits a certain level of robustness to noise. When comparing different scaling factors for each dataset, we observe that the scaling factor of $8\times$ is relatively more robust to the increasing level of noise.}

\clr{
The results demonstrate that in the presence of noise, {\sysname} still maintains its effectiveness in reconstructing the high-resolution data to some extent. 
The primary focus of this paper is not on noisy data, however, this is an important point to consider in future work. To improve the model's robustness to noises, one should consider introducing noises during training, so that during testing, the model knows how to reconstruct the clean data from noisy input.}

\begin{figure}[htp]
    \centering
    \includegraphics[width=\columnwidth]{figures/noisy_low_res.pdf}
    \vspace{-10pt}
    \caption{\clr{{\sysname}'s super-resolution PSNRs for low-resolution input with different noise levels.}}
    \label{fig:nosiy_input}
\vspace{-6pt}
\end{figure}

\section{\clr{Model Size and Training Time}}
\clr{We have created a table (\cref{tab:train_time}) displaying the training time for each model and each dataset. Note that the model sizes are the same for different datasets, this information is shown in Table 2 of the main text.}

\section{\clr{Summary of Notations}}
\clr{\cref{tab:symbol} is a summary of concepts and notations in the paper.}


\begin{table}[!ht]
\small
\caption{\clr{Dataset name and total training time for each dataset.}}
\centering
 \begin{tabular}{c|c c}
 Dataset & Method & Train Time \\ [0.5ex] 
  \hline
  \multirow{4}*{Vortex} & \makecell{SSR-TVD} & \makecell{23h26m} \\
  ~ & \makecell{SSR-TVD (w/o D)} & \makecell{23h4m} \\
  ~ & \makecell{ESRGAN} & \makecell{23h43m} \\
  ~ & \makecell{{\sysname} (our)} & \makecell{\textbf{23h48m}} \\
  \cline{1-3}

  \multirow{4}*{Nyx} & \makecell{SSR-TVD} & \makecell{16h41m} \\
  ~ & \makecell{SSR-TVD (w/o D)} & \makecell{16h56m} \\
  ~ & \makecell{ESRGAN} & \makecell{23h58m} \\
  ~ & \makecell{{\sysname} (our)} & \makecell{\textbf{23h32m}} \\
  \cline{1-3}

  \multirow{4}*{Combustion} & \makecell{SSR-TVD} & \makecell{22h48m} \\
  ~ & \makecell{SSR-TVD (w/o D)} & \makecell{22h27m} \\
  ~ & \makecell{ESRGAN} & \makecell{23h19m} \\
  ~ & \makecell{{\sysname} (our)} & \makecell{\textbf{23h6m}} \\
  \cline{1-3}

  \multirow{4}*{Plume} & \makecell{SSR-TVD} & \makecell{23h10m} \\
  ~ & \makecell{SSR-TVD (w/o D)} & \makecell{23h24m} \\
  ~ & \makecell{ESRGAN} & \makecell{23h28m} \\
  ~ & \makecell{{\sysname} (our)} & \makecell{\textbf{23h18m}} \\
  \cline{1-3}
 \end{tabular}
 \label{tab:train_time}
\end{table}

\begin{table}[!ht]
\small
\caption{\clr{Symbol name and meaning.}}
\centering
 \begin{tabular}{c|c} 
 Symbol & Meaning \\ [0.5ex] 
 \hline
 CNF & conditional normalizing flow \\
 CNF$^{-1}$ & inverse of conditional normalizing flow \\
 LR & low-resolution \\
 HR & high-resolution \\
 $X$ & random variable (high-resolution data) with distribution $P(X)$\\
 $Y$ & random variable (low-resolution data) with distribution $P(Y)$ \\ 
 $Z$ & latent variable with distribution $P(Z)$\\
 r & upscaling factor for each data dimension\\
 $D,H,W$ & $D$: Depth, $H$: Height, $W$: Width, of low-resolution data\\
 $\mathbf{x}$ & an instance of random variable $X$ (input to the normalizing direction of flow g)\\ 
 $\mathbf{y}$ & an instance of random variable $Y$ (input to the encoder e)\\
 $\mathbf{z}$ & an instance of random variable $Z$\\
 $p(\mathbf{x})$, $p(\mathbf{y})$, $p(\mathbf{z})$ & probability density of $\mathbf{x}$, $\mathbf{y}$ and $\mathbf{z}$ \\ 
 S, T & neural networks that predict scale and translation parameters\\ 
 g & unconditional normalizing flow \\
 h & conditional normalizing flow \\
 g$^{-1}$ & inverse of unconditional normalizing flow g \\
 h$^{-1}$ & inverse of conditional normalizing flow h \\
 e & ResNet encoder \\
 $\mathbf{z}_h$, $\mathbf{z}_l$ &  high- and low-frequency latent representations \\
 $\mathbf{z}_{lat}$ & low-resolution latent representation, output of encoder e\\
 $\mathbf{z}_{0}$ & the innermost Gaussian latent representation \\
 $m_{\mu}$, $m_{\Sigma}$ & neural networks that predict Gaussian distribution parameters for $\mathbf{z}_{0}$ \\
 $\mathcal{N}(\mu,\Sigma)$ & Gaussian distribution with mean vector $\mu$ and covariance matrix $\Sigma$\\
 \end{tabular}
 \label{tab:symbol}
\end{table}

\begin{figure}[htp]
    \centering
    \includegraphics[width=0.7\columnwidth]{figures/flow_step.pdf}
    \vspace{-6pt}
    \caption{\clr{Each flow step in the normalizing flow model contains an actnorm layer, an invertible $1 \times 1 \times 1$ convolution layer, and a coupling layer.}}
    \label{fig:flowstep}
\vspace{-8pt}
\end{figure}

\begin{figure}[t]
    \centering
    \includegraphics[width=\columnwidth]{figures/modelsize_N.pdf}
    \vspace{-18pt}
    \caption{\clr{Top row: PSNR for the different number of flow steps and different upscale factors for two ensemble members of the Nyx data. Bottom row: PSNR and standard deviation (std) of uncertainty values for Vortex data given different N with upscale factors $2\times$ and $4\times$.}}
    \label{fig:num_layers_n}
\vspace{-8pt}
\end{figure}

\begin{figure*}[htp]
    \centering
    \includegraphics[width=0.9\textwidth]{figures/vortex_all_v2.pdf}
    \caption{Volume rendering images of baselines' and {\sysname}'s super-resolution results, ground truth (GT), and low-resolution input (Low Res) for the Vortex data.}
    \label{fig:vortex_all}
\end{figure*}

\section{\clr{Flow Architecture}}
\clr{
As illustrated in \cref{fig:flowstep}, each flow step includes three basic invertible transformations as discussed in Section 3 in the main text: 
(1) an activation normalization layer (actnorm) that performs channel-wise normalization on feature maps to stabilize training and mitigates gradient exploding and gradient vanishing problems, (2) an invertible $1 \times 1 \times 1$ convolution layer, which is a generalization of the channel permutation operation, and (3) an affine coupling layer that contains complex transformations to model complicated dependencies between variables. 
By stacking multiple flow steps, the model has the ability to learn the transformations between the Gaussian latent space and the high-resolution data space. }

\section{\clr{Hyperparameter Analysis}}
\clr{We study the influence of the number of flow steps on performance, and the number of samples for uncertainty estimation in this section. }
\clr{
\textbf{Evaluation of the number of normalizing flow steps.} {\sysname} consists of two normalizing flow components, i.e., a standard normalizing flow $g$ followed by a conditional normalizing flow $h$. We evaluate how the number of flow steps influences the performance by training five different models with varying numbers of flow steps on the same training data.
\Cref{fig:num_layers_n} (top) shows bar charts comparing the PSNR values of reconstructed high-resolution results for these models on two ensemble members of Nyx data. The x-axis in the bar chart represents the model size in format $[K_g,K_h]$, where $K_g$ and $K_h$ denote the number flow steps for $g$ and $h$, respectively. The y-axis represents the PSNR of the high-resolution output. Colors in the plot correspond to different upscale factors. By comparing the PSNR values of these models for both ensemble members, we find that the number of flow steps will affect the model's performance. Shallow models produce low-quality results since it is difficult to transform between a complicated data distribution and a Gaussian latent distribution with a small number of flow steps. However, when the model is very deep, it can be difficult to converge, and adding more flow steps will not improve the performance but only slow down the computation. We use slightly different architecture settings for each dataset, but unless using extreme settings, the performance variation is small. In our experiments, for the Vortex and Nyx data, we set $K_g=5$ and $K_h=5$. For the Combustion and Plume data, we set $K_g=3$ and $K_h=3$.}

\clr{
\textbf{Evaluation of the number of samples for uncertainty estimation.} For uncertainty quantification, we need to sample the Gaussian latent space $N$ times and compute the super-resolution outputs' variance. However, the computation time will increase as we increase the number of samples for a more accurate estimation of variance.
$N$ can be heuristically chosen \cite{hogg1977probability} for an unbiased estimation of the distribution's variance. To find an optimal $N$, in~\cref{fig:num_layers_n} (bottom), we plot PSNR values of super-resolved outputs and the standard variations of uncertainty values for different values of $N$, with upscale factors $2\times$ and $4\times$, respectively. In our experiments, we take $N=40$ for uncertainty estimation when PSNR and the standard variation become stable. 
}

\section{Additional Super-Resolution Rendering Images}
In this section, we present volume rendering images of the low-resolution inputs and their corresponding super-resolution results for all datasets with scale factors of $2\times$, $4\times$, and $8\times$, as shown in \cref{fig:vortex_all}, \cref{fig:nyx_all}, \cref{fig:combustion_all}, and \cref{fig:plume_all} in this supplementary material. We also present isosurface rendering images for the Vortex data with $isovalue=6$, as shown in \cref{fig:vortex_iso_all}. 
Results for the same dataset are using the same transfer function. 
The result also suggests that a future direction to improve the super-resolution performance is to enhance the quality of the saved low-resolution data.

\begin{figure*}[htp]
    \centering
    \includegraphics[width=\textwidth]{figures/vortex_all_iso_v2.pdf}
    \caption{Isosurface rendering images of baselines' and {\sysname}'s super-resolution results, ground truth (GT), and low-resolution input (Low Res) for the Vortex data. 
    Due to the significant information loss in low-resolution data, reconstructing the exact features in high-resolution data is challenging. While some missing features can still be reconstructed by learning-based methods but is impossible for the interpolation-based method.}
    \label{fig:vortex_iso_all} 
\end{figure*}

\begin{figure*}[htp]
    \centering
    \includegraphics[width=\textwidth]{figures/nyx_all_v2.pdf}
    \caption{Volume rendering images of baselines' and {\sysname}'s super-resolution results, ground truth (GT), and low-resolution input (Low Res) for the Nyx data.}
    \label{fig:nyx_all}
\end{figure*}

\begin{figure*}[htp]
    \centering
    \includegraphics[width=\textwidth]{figures/combustion_all_v2.pdf}
    \caption{Volume rendering images of baselines' and {\sysname}'s super-resolution results, ground truth (GT), and low-resolution input (Low Res) for the Combustion data.}
    \label{fig:combustion_all}
\end{figure*}

\begin{figure*}[htp]
    \centering
    \includegraphics[width=0.7\textwidth]{figures/plume_all_v2.pdf}
    \caption{Volume rendering images of baselines' and {\sysname}'s super-resolution results, ground truth (GT), and low-resolution input (Low Res) for the Plume data.}
    \label{fig:plume_all}
\end{figure*}

\bibliographystyle{abbrv-doi-hyperref}

\bibliography{template}

\appendix 